\documentclass[aps,prc,twocolumn,showpacs,preprintnumbers,
               longbibliography,nofootinbib,float,superscriptaddress]{revtex4-1}
               
\usepackage{color}
\usepackage[dvipsnames]{xcolor}
\usepackage{float,amsmath}
\usepackage{graphicx}
\usepackage[utf8]{inputenc}
\usepackage{hyperref}

\newcommand{\xt}{{\mathbf{x}_\perp}}

\def\st{\begin{equation}}
\def\stp{\end{equation}}
\def\Eq#1{Eq.~(\ref{#1})}

\def\Fig#1{Fig.~\ref{#1}}

\def\Sect#1{Sect.~\ref{#1}}
\def\Ref#1{Ref.~\cite{#1}}

\def\llangle{\left\langle}
\def\rrangle{\right\rangle}

\def\ptbar{{\bar p_T} }
\def\deltan{\hat\delta}
\def\NN{\mathcal N}
\def\sigman{\hat\sigma}
\newcommand{\ave}[1]{\llangle #1 \rrangle }
\newcommand{\nnbar}[1]{ \ave{\NN(#1)}}

\begin{document}

\title{Transverse momentum fluctuations and their correlation with elliptic flow\\ in nuclear collisions}

\author{Bj\"orn Schenke}
\affiliation{Physics Department, Brookhaven National Laboratory, Upton, NY 11973, USA}

\author{Chun Shen}
\affiliation{Department of Physics and Astronomy, Wayne State University, Detroit, MI 48201, USA}
\affiliation{RIKEN BNL Research Center, Brookhaven National Laboratory, Upton, NY 11973, USA}

\author{Derek Teaney}
\affiliation{Department of Physics and Astronomy, Stony Brook University, Stony Brook, NY 11794, USA}

\begin{abstract}
   We propose observables $v_0$ and $v_0(p_T)$ which quantify the relative
   fluctuations in the total transverse momentum at fixed multiplicity. We
   first study the factorization of the fixed multiplicity 
  momentum dependent 
   two particle correlation function into a product of $v_0(p_T^a)$ and $v_0(p_T^b)$ within
   realistic hydrodynamic simulations. Then we present computations of $v_0(p_T)$
   for different particle types.
   We determine the relation between the integrated $v_0$ and previously measured
   observables, and compare results from a hybrid hydrodynamics based model to
   experimental data. The effects of bulk viscosity and an initial
   pre-equilibrium stage on the results are quantified. 
   We find that $v_0$ is strongly correlated with the initial state entropy per elliptic area, $S/A$.
 Using this result, 
 we explain how the observed correlations between the elliptic flow and 
 the transverse momentum (both in simulations and experiment) reflect the initial
 state correlations between $1/A$ and ellipticity $\varepsilon_2$ at fixed multiplicity.
We argue that the systematic experimental study of $v_0$, with the same sophistication as used for the other $v_n$, can contribute significantly to our understanding of quark gluon plasma properties.
\end{abstract}

%\pacs{11.15Bt, 04.25.Nx, 11.10Wx, 12.38Mh}
\maketitle

\section{Introduction}

One of the goals of the experimental program of heavy ion collisions at the Relativistic Heavy Ion Collider (RHIC) and the Large Hadron Collider (LHC) is to quantify the properties of the Quark Gluon Plasma (QGP)  by measuring the response of the nuclear medium to gradients in the energy
density.  If the system is close enough to equilibrium, this response can be
characterized by hydrodynamics. Indeed, hydrodynamic
simulations,  based on approximate thermal equilibrium, describe an enormous variety of data on the long range azimuthal correlations amongst the produced hadrons~\cite{Teaney:2009qa,Gale:2013da,Luzum:2013yya,Heinz:2013th,Jeon:2015dfa}.  Hydrodynamic fits
to experimental results on elliptic and triangular flows (as well as other observables) place increasingly precise constraints on the shear and bulk viscosity of the QGP as well as its equation of state~\cite{Bazavov:2014pvz,Moreland:2018gsh,Gardim:2019brr}. 

Most of the experimental and theoretical analyses  have focused on the harmonic spectrum. In this case, the azimuthal distribution of particles in each event is expanded in a Fourier series
\st
 2\pi \frac{dN}{d\phi} = N  \left[ 1 + \sum_{n=1}^{\infty} V_n e^{in \phi}  + {\rm c.c. } \right]\,,
\stp
where $V_n$ is a complex Fourier coefficient and ${\rm c.c.}$ denotes the complex conjugate. The harmonic spectrum is 
further binned in transverse momentum\footnote{Throughout the paper the transverse momentum is denoted $p\equiv p_T$, i.e., the $\scriptsize{T}$ subscript will sometimes be dropped to lighten the notation. In addition we will generally assume boost invariance, so that $N$ is short for $dN/d\eta$ etc.}
\st
\label{vnp}
 2\pi \frac{dN}{dp\, d\phi} = \frac{dN}{dp} \left[ 1 +  \sum_{n=1}^{\infty} V_n(p) e^{in \phi}  + {\rm c.c. } \right] \, .
\stp
The  physical origin of 
the fluctuations in $V_n$  and $V_n(p)$ is the following:
there are event-by-event fluctuations in the angular distribution of the initial energy density, these fluctuations (and their gradients)  drive the collective flow of the system, and this collective flow  is ultimately  imprinted on the particles and their associated harmonic spectrum, $V_n(p)$. 
In practice, it is the long-range 
correlations (in pseudo-rapidity) between the produced particles that are measured \cite{Luzum:2013yya,Heinz:2013th}.
The two-particle correlation function yields precise data for the squared Fourier coefficients
\st
v_n^2 \equiv \llangle |V_n|^2 \rrangle  \, ,
\stp
and the corresponding momentum dependent measures
\st
v_n(p) \equiv  \frac{\llangle V_{n}(p)  V_n^* \rrangle }
{ \sqrt{\llangle |V_n|^2 \rrangle } } \, .
\stp
Here the angular brackets denotes an average over events.
A common scheme to eliminate short range correlations is to require that the correlated pairs are separated by a rapidity 
gap of order $\Delta \eta > 2$.
The study of flow gives direct information on the medium response to the
energy gradients in the system  and the fluctuations in the initial state.  

In traditional flow analyses,  fluctuations in the transverse momentum, i.e., the event-to-event distribution of $dN/dp$  in \Eq{vnp}, are either
not considered or described in somewhat different terms than the $v_n$~\cite{PhysRevC.72.044902,Abelev:2014ckr}.   
This distribution is characterized by the two point functions, $\llangle dN/dp_1 \, dN/dp_2 \rrangle$. 
Except in the the principal component analysis of \cite{Sirunyan:2017gyb},
available measurements on momentum fluctuations~\cite{PhysRevC.72.044902,Abelev:2014ckr,Aad:2019fgl}   
have neglected the rapidity gap typical in flow analyses, and thus  reflect both non-flow correlations and long-range hydrodynamic fluctuations,  obscuring the underlying physics.

In a hydrodynamic picture, the physical origin of the
long range momentum correlations is identical to the $v_n$ case~\cite{Bozek:2012fw,Bozek:2016yoj}:  there
are event-to-event fluctuations in the radius (at fixed multiplicity), these
size fluctuations lead to fluctuations in the radial flow, and this flow is in turn 
imprinted on the momentum spectra of the produced particles.  
Although momentum fluctuations are perhaps more sensitive to the equation of state and the bulk viscosity of the QCD plasma than harmonic flow~\cite{Ryu:2015vwa,Gardim:2019xjs}, momentum fluctuations are not as well studied.

Building on several papers which will be discussed more completely in what
follows~\cite{Bhalerao:2014mua,Olszewski:2017vyg,Gardim:2019iah}, we will define an observable $v_0(p)$ (analogous to
$v_n(p)$) that quantifies the momentum fluctuations and is
straightforward to measure and interpret. The only minor complication in defining
$v_0(p)$ is that the multiplicity should be held fixed. The partial
covariance method  is the right tool for this job~\cite{Olszewski:2017vyg}, and
provides a simpler alternative to the ``centrality bin width
correction''~\cite{[{see for example, }] PhysRevLett.103.172301} and to principal component
analysis~\cite{Bhalerao:2014mua,Mazeliauskas:2015efa,Gardim:2019iah,Hippert:2019swu}.
The principal component decomposition can be sensitive to non-flow and lacks a compelling measure~\cite{Liu:2020ely}. 
We hope that measurements of $v_0(p)$ for different particles
and perhaps jets will become standard. Such measurements 
can provide complementary information to the harmonic spectrum,  
shedding
light on thermalization in small systems,
and the poorly understood ``no-man's'' land, a region of $p_T$ from $2 \ldots 6\,{\rm GeV}$.
Inspired by earlier work, we have found that the fluctuation in the entropy per  area is an excellent predictor for the $v_0(p)$, and thus measurements of $v_0(p)$ can 
be used to constrain this important property of the initial state. The integrated quantity $v_0$ is closely
related to early measurements of momentum fluctuations quantified with the nondescript variable, $C_m$~\cite{PhysRevC.72.044902,Abelev:2014ckr,Aad:2019fgl}. 
In \Sect{sec:Cm} we will determine the relation between $C_m$ and the momentum integrated $v_0$. We will also discuss the relation between $v_0(p)$ and principal components in \Sect{sec:principal}, borrowing heavily from the discussion in \Ref{Gardim:2019iah}.

In \Sect{sec:results}, we present hydrodynamic simulations with the \textsc{Music} ~\cite{Schenke:2010nt,Schenke:2010rr,Schenke:2011bn,Paquet:2015lta}+UrQMD \cite{Bass:1998ca,Bleicher:1999xi} hybrid model with IP-Glasma initial conditions~\cite{Schenke:2012wb,Schenke:2012hg}. We show results for the fixed multiplicity two particle correlation function that measures correlations of the transverse momentum spectrum at different $p_T$, and can be used to define $v_0(p_T)$ under the assumption of factorization. We show the quality of the factorization in the same section, and compare our results to a simple model from \Ref{Gardim:2019iah} for the fluctuations of the transverse momentum spectrum. We then present results for $v_0(p_T)$ for different particle species and various centralities, and provide quality measurements for different estimators of transverse momentum fluctuations.

In \Sect{Sect:ComparisonWpt} we present comparisons of the hybrid model calculations with existing data on transverse momentum fluctuations, and discuss the effects of bulk viscosity and a pre-equilibrium stage described by the KoMPoST model \cite{Kurkela:2018vqr,Kurkela:2018wud}.

Recently, the ATLAS Collaboration, motivated by 
earlier theoretical work~\cite{Mazeliauskas:2015efa,Bozek:2016yoj},  
has studied the correlations between $v_2$ and $v_3$ and the transverse momentum fluctuations~\cite{Aad:2019fgl}. 
We will compare the hybrid model simulations described above to this recent data in \Sect{sec:v2pTcorrelations}.  In addition, we will present estimators for this observable, constructed from initial state properties. We find that the entropy per (elliptic) area provides a good initial state predictor for the mean transverse momentum fluctuations $\delta\bar{p}_T$, which together with the eccentricity as estimator for the elliptic flow, estimates the $v_2^2$-$\bar{p}_T$ correlations and their centrality dependence reasonably well. 
We further show the predictor obtained from a Monte-Carlo Glauber model calculation with high statistics, which also reproduces the characteristic features of the experimental data, reaffirming their geometric origin.

We conclude in \Sect{sec:conclusions} and emphasize again the power of detailed $v_0(p_T)$ measurements to shed light on quark gluon plasma properties, including thermalization in small systems and the onset of jet quenching.

While  finalizing this manuscript a paper by Bozek and Mehrabpour  appeared, which partially overlaps with the current work both in methodology and conclusions~\cite{Bozek:2020drh}.  We will note the similarities and differences with this paper below.

\section{Momentum fluctuations and $v_0$} 
\label{sec:v0defs}

For any observable $O$, the event-by-event deviation
and variance are defined as 
\st
\delta O \equiv O - \llangle O \rrangle \, , \qquad \sigma_O^2 = \langle (\delta O)^2  \rangle \,  .
\stp
Similarly, the 
event-by-event deviation and variance at  \emph{fixed} multiplicity are 
defined as~\cite{Olszewski:2017vyg} 
\begin{align}
\deltan O &\equiv \delta O  - \frac{\llangle \delta O \delta N \rrangle }{ \sigma_N^2 } \delta N \,, \\
\sigman_O^2 &\equiv \langle (\deltan O)^2 \rangle  = \llangle (\delta O)^2 \rrangle - \frac{\llangle \delta O \delta N \rrangle^2 }{\sigma_N^2 } \, .
\end{align}
It follows that the covariance between two observables at fixed multiplicity is
\st
\label{eq:O1O2} 
\llangle \deltan O_1 \deltan O_2 \rrangle =  \llangle \delta O_1 \delta O_2 \rrangle - \frac{ \llangle \delta O_1 \delta N \rrangle \llangle \delta O_2 \delta N \rrangle  }{ \sigma_N^2 } \,.
\stp
The subtraction terms are designed to remove the linear correlation between the observable and the multiplicity.  Below 
we will use this so-called ``partial correlation method''  to analyze the fluctuations of transverse momentum and elliptic flow at fixed multiplicity. This approach was recently used in \Ref{Bozek:2020drh} for the same purpose. 

It is important to emphasize that the primary goal of measuring the variance at fixed multiplicity is to remove trivial centrality fluctuations. While we have used the multiplicity as a centrality measure, other quantities, such as the forward transverse energy, can be used. 
In the ATLAS detector for instance, a natural choice would be the calorimetric measure FCal $\mathcal E \equiv \sum E_T$, leading to the definition
$\deltan O= \delta O - \llangle \delta O \delta \mathcal E \rrangle \delta \mathcal E/\sigma_{\mathcal E}^2$. This definition would leave the central tracker free to make the correlation measurement in \Eq{eq:O1O2}. In a different context, two separate correlation measurements, one at fixed forward $\mathcal E$ and one at fixed multiplicity, have already been performed, and the two measurements give nearly identical results, see the right two panels of Fig.\,1 from \Ref{ATLAS:2015kla}.   

The event-by-event spectra  and multiplicity $N$ are notated
\[ \NN  (p) \equiv \frac{dN}{dp} \,, \qquad  N = \int_0^{\infty} dp \, \NN(p) \, , 
\]
We are interested in the fluctuations in the spectra at fixed multiplicity:
\st
\label{fluctatfixed}
\deltan \NN(p^a) = \delta \NN(p^a) -  \frac{\llangle \delta \NN(p^a) \delta N \rrangle  }{\sigma_N^2 }  \delta N \, .
\stp
 The fluctuations in integrated $p_T$ at fixed multiplicity
can be used to characterize 
the spectral fluctuations, and thus we define
\begin{subequations}\label{eq:deltahatpT} 
\begin{align}
   P_T \equiv& \int_0^\infty dp  \, p \, \NN(p) \,,  \\
\deltan P_T =& \int_0^\infty dp  \, p \, \deltan \NN(p)  \, .
\end{align} 
\end{subequations}
$\deltan P_T$ is analogous to the $\vec{Q}_2$-vector used to define the elliptic flow \cite{Ollitrault:1992bk,Voloshin:1994mz,Barrette:1994xr}. Specifically it is a \emph{sum} over particles and therefore characterizes the collective response.
We define the integrated $v_0$ via the variance in the integrated $P_T$ at fixed multiplicity
\st
\label{v02def}
v_0^2 =  \frac{ \sigman_{P_T}^2 }{\llangle P_T \rrangle^2 }  \, .
\stp
The momentum dependent $v_0(p)$ can be defined as:
\st
\label{oneparticledef}
v_0(p) = \frac{1}{\ave{\NN(p)}}\,  \frac{\llangle \deltan \NN(p) \, \deltan P_T \rrangle }{ \sigman_{P_T} } \, .
\stp
It is evident from Eqs.~(\ref{eq:deltahatpT}), (\ref{v02def}), and (\ref{oneparticledef})
that the integrated $v_0$ is
determined by $v_0(p)$ according to
\st
v_0 \equiv \frac{\int_0^{\infty} dp \, p \, \ave{\NN (p)} \, v_0 (p)   }{
\int_0^{\infty} dp \, p \, \ave{\NN (p)}  }\,.
\stp
We also note that using \Eq{fluctatfixed} it follows that
\begin{equation}
   \label{integratedquantity}
   \int_0^{\infty} dp \, \nnbar{p} \, v_0 (p) = 0  \, .
\end{equation}
More generally, if the forward calorimeter is used to define centrality,
then the zero on the r.h.s. of \Eq{integratedquantity} is replaced with 
the variance of the  multiplicity at fixed forward energy, $\sigma_N^2 - \llangle \delta N \delta \mathcal E \rrangle^2/\sigma_{\mathcal E}^2$, which is an interesting quantity in its own right. 

It is also interesting to measure $v_0(p)$ for a variety of particles,
such as $D$ mesons and $J/\psi$, and perhaps jets.   We define for species $s$ 
\st
\label{eq:v0speciesdef}
v_{0,s} (p) = \frac{1}{\ave{\NN_s(p)}}\,  \frac{\llangle \deltan \NN_s(p) \, \deltan P_T \rrangle }{ \sigman_{P_T} }\,,
\stp
where the quantities without index $s$ are obtained using all charged hadrons.

 In practice, when assuming factorization, $v_0(p)$ can be obtained from two particle correlations. Generally, the two particle correlation function is given by
\begin{align}
\label{NNDelta}
    {\mathcal N}_{ \Delta}(p^a, p^b)   &\equiv 
    \llangle \delta \NN(p^a) \delta \NN(p^b) \rrangle
    \notag\\ &= 
    \llangle  \NN(p^a) \NN(p^b) \rrangle - \llangle  \NN(p^a) \rrangle \llangle \NN(p^b) \rrangle \, ,
\end{align}
and the correlation function of deviations from the event averaged $\langle \NN(p)\rangle$ at fixed multiplicity is
\begin{align}
    &\llangle \deltan \NN(p^a) \deltan \NN(p^b) \rrangle \equiv \notag\\
    &  ~~ \llangle \delta\NN(p^a) \delta \NN(p^b) \rrangle  - \frac{\llangle \delta \NN(p^a) \delta N \rrangle \llangle \delta \NN(p^b) \delta N \rrangle }{\sigma_N^2 } \, .
\end{align}
Now we divide by the mean spectra and define $v_0(p)$ from
the two particle correlation function
\st
\label{twoparticledef}
C(p^a,p^b) = \frac{\llangle \deltan \NN(p^a) \deltan \NN(p^b) \rrangle }{\ave{ \NN(p^a) }  \, \ave{ \NN(p^b) } } \approx v_0(p^a) v_0(p^b).
\stp
Here we have assumed that the spectrum of fluctuations factorizes into a product of a function of $p^a$ and a function of $p^b$.   
This will need to be checked experimentally. 
Provided this factorization holds, one can integrate over $p^b$ on both sides of \Eq{twoparticledef}, $\int dp^b p^b \nnbar{p^b} \ldots$, and verify the consistence with \Eq{oneparticledef}.  

In practice,  to reduce non-flow
the two particle correlations probed by $v_0(p_T)$ 
should be computed using a rapidity gap of approximately two units. When a gap is used these formulas should be modified
appropriately.
We also  note that all efficiencies cancel as in the $v_2(p_T)$ case, making for a straightforward measurement.

\subsection{Relation of $v_0(p)$ to slope fluctuations in a simple model}
\label{sec:OlliModel}

Now we determine how $v_0(p)$ is related to event-by-event slope modifications of the spectra. To this end, we recall the analysis and  model of \Ref{Gardim:2019iah}, which studied principal components of $\NN(p)$. 
In this model the spectrum is written as
\st\label{eq:modelspectrum}
\NN(p_T) =  (2\pi p_T)\,  N  \frac{ e^{-2 p_T/\ptbar}  }{\pi \ptbar^2 } \, , 
\stp
where the parameters  $N$ and $\ptbar$ fluctuate from event to event.
Here $N$ is the event-by-event multiplicity and  $\ptbar$ is the 
event-by-event mean $p_T$, while the leading factor $(2\pi p_T)$ is the appropriate measure.
Thus, if the parameters $N$ and $\ptbar$ fluctuate, the spectrum fluctuates
as
\st
  \frac{\delta \NN(p_T)}{\llangle \NN(p_T) \rrangle }  
  = \frac{\delta N}{\llangle N \rrangle } - \frac{2 \delta \ptbar }{\llangle \ptbar \rrangle } + 2\frac{ p_T \delta \ptbar   }{\llangle \ptbar \rrangle ^2 } \, .
\stp
Then, unraveling the nested definitions, one finds after a certain amount of
algebra
\st \label{eq:modelresult}
     \frac{\llangle \deltan \NN(p_T^a) \, \deltan \NN(p_T^b) \rrangle }{
        \nnbar{p_T^a} \, \nnbar{p_T^b} 
 } =  \frac{\sigman_{\bar p_T}^2}{\llangle \ptbar \rrangle^2}  
 \left(\frac{2 p_T^a}{ \ave{\ptbar} } - 2   \right)\left(\frac{2 p_T^b}{ \ave{\ptbar}} - 2   \right) \, , 
\stp
where we recall that $\sigman_{\ptbar}^2  =  \sigma_{\ptbar}^2   - \llangle \delta \ptbar \delta N\rrangle^2 /\sigma_N^2$. It follows for this model that
\st
\label{eq:v0ptmodel}
v_0(p_T) = \frac{\sigman_{\bar p_T}}{\llangle \ptbar \rrangle}  \left(\frac{2 p_T}{ \ave{\ptbar}} - 2   \right) \, .
\stp
As could be anticipated, $v_0(p_T)$ describes the fluctuations in the event-by-event mean $\ptbar$ at fixed multiplicity and increases linearly with $p_T$. 
We will determine in \Sect{sec:results} how well
this simple model describes the correlator $C(p^a,p^b)$ computed within
the IP-Glasma+\textsc{Music}+UrQMD hybrid model.

\subsection{Relation of $v_0$ to previous measurements}
\label{sect:relationto}

In this section we compare the integrated $v_0$ to other measures of $p_T$ fluctuations which exist in the literature.

\subsubsection{$C_m$} \label{sec:Cm}
The ALICE collaboration defines a variance $C_m$~\cite{Abelev:2014ckr}, which is slightly different from the definition used by ATLAS~\cite{Aad:2019fgl}.  In the notation used above one defines
\st
\Delta P_T = \delta P_T - \frac{\llangle P_T \rrangle  }{\llangle N \rrangle } \delta N \, ,
\stp
and\footnote{In practice, self correlations are excluded in these averages.} 
\st
C_{m} = \frac{\llangle \Delta P_T \Delta P_{T} \rrangle }{ \llangle N^2 \rrangle } \, . 
\stp
Traditionally, because $p_T$ fluctuations were considered distinct from flow measurements,  no rapidity gap  was used.

These measurements suffer from non-flow and unfortunately can not
be fairly compared to most hydrodynamic simulations. We will ignore
the issue of non-flow in this work, and directly compare to the data, while we eagerly await a modern measurement of radial flow fluctuations.

The ALICE Collaboration defines a mean $p_T$ measure $M(p_T)$ :
\begin{subequations}
   \label{Mofpt}
\begin{align}
   \label{Mofpta}
M(p_T) &\equiv \frac{\llangle P_T \rrangle}{\ave{N}} = \frac{\llangle \ptbar N \rrangle}{\ave{N}} = \llangle \ptbar \rrangle   +  \frac{\llangle \ptbar \delta N \rrangle }{\ave{N} } \, , \\
   \label{Mofptb}
       & \simeq   \ave{\ptbar} \left(1   + \mathcal O\left(\frac{1}{\ptbar} \frac{d\ptbar}{dN} \, \frac{\sigma_{N}^2}{\llangle N \rrangle} \right) \right) \, .
\end{align}
\end{subequations}
If the  multiplicity bins are very narrow (as was the case for the ALICE
measurements~\cite{Abelev:2014ckr}), then the fluctuations in $N$ are small and the second term in \Eq{Mofpta} can be neglected. In this case  $M(p_T)$ is approximately $\llangle \ptbar\rrangle$.
The partial correlation adopted in this work removes
any sensitivity to the bin width, and larger bins  (with higher statistics) can be used.

With the same approximations, $C_m$ is  the variance in  $\ptbar$.
Writing $P_T = \ptbar N$,  and substituting 
$\ptbar  = \llangle \ptbar \rrangle + \delta\ptbar$ and $N = \llangle N \rrangle +\delta N$, yields
\begin{subequations}
\begin{align}
   \label{eq:DPt1}
   \frac{\Delta P_T}
   { \llangle \ptbar \rrangle \llangle N \rrangle}  
   =&  
   \frac{\delta \ptbar}{\llangle \ptbar \rrangle}   - 
   \frac{\llangle \delta \ptbar \delta N \rrangle} {\llangle \ptbar  \rrangle \llangle N \rrangle^2 } \delta N  \, ,  \\
   \label{eq:DPt2}
   \simeq & 
   \frac{\delta \ptbar}{\llangle \ptbar \rrangle} \, . 
\end{align}
\end{subequations}
The last term in \Eq{eq:DPt1} is  small (see \Eq{Mofpt} and surrounding text), leading to the expected result 
\st \label{eq:CmApprox}
\frac{  C_{m}} {M(p_T)^2 } \simeq  \frac{\sigma_{\ptbar}^2 }{\ave{\ptbar}^2 } \, .
\stp
Finally, the ATLAS Collaboration defines a slightly different quantity from $C_m$, called $c_k$
which is closely related to $C_m$ but studies the variance in $\delta (P_T/N)$. We will not go 
through the details here, but within the same approximations,   $C_m \simeq c_k$.

\begin{figure*}[t]
\includegraphics[width=0.48\textwidth]{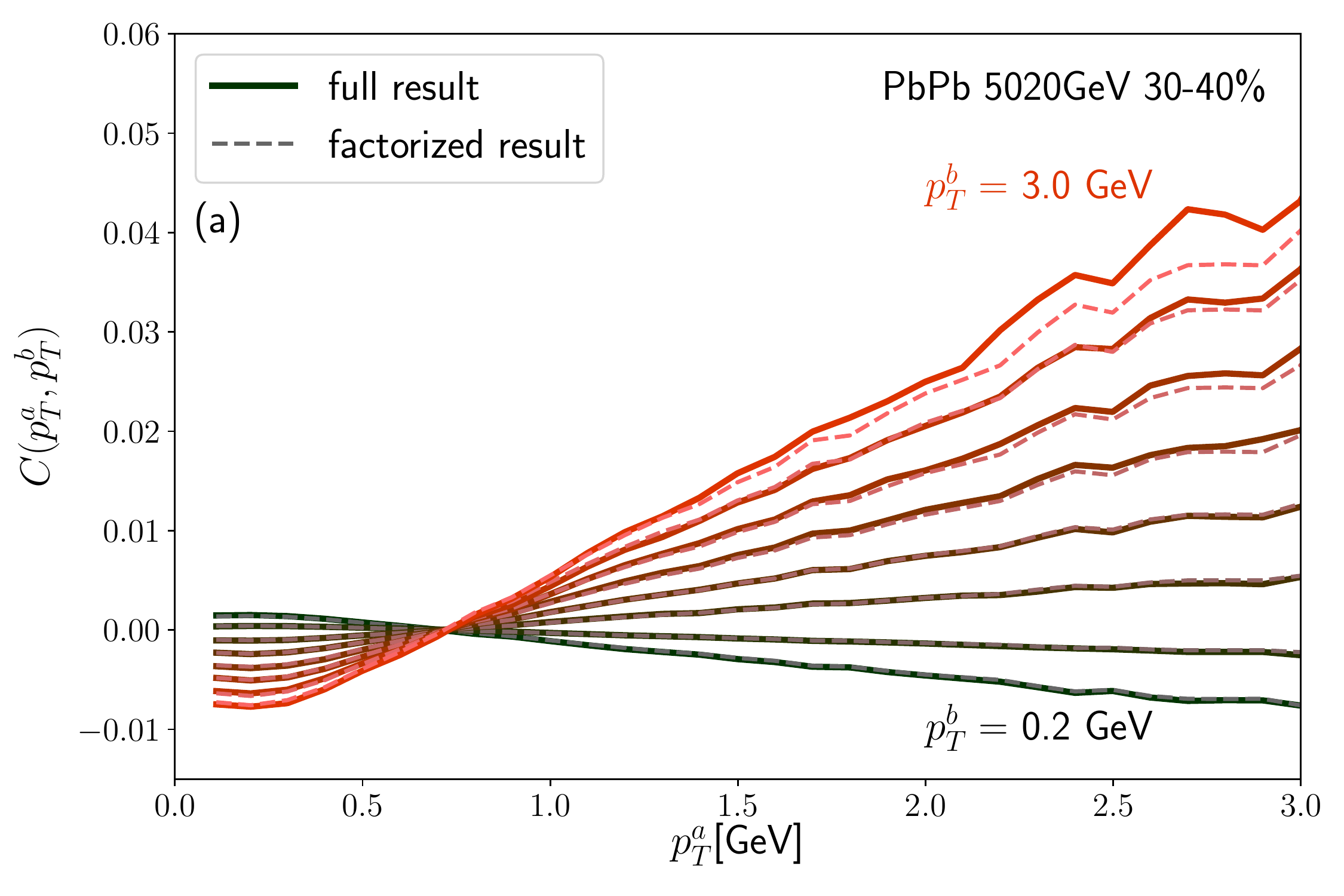}
\includegraphics[width=0.48\textwidth]{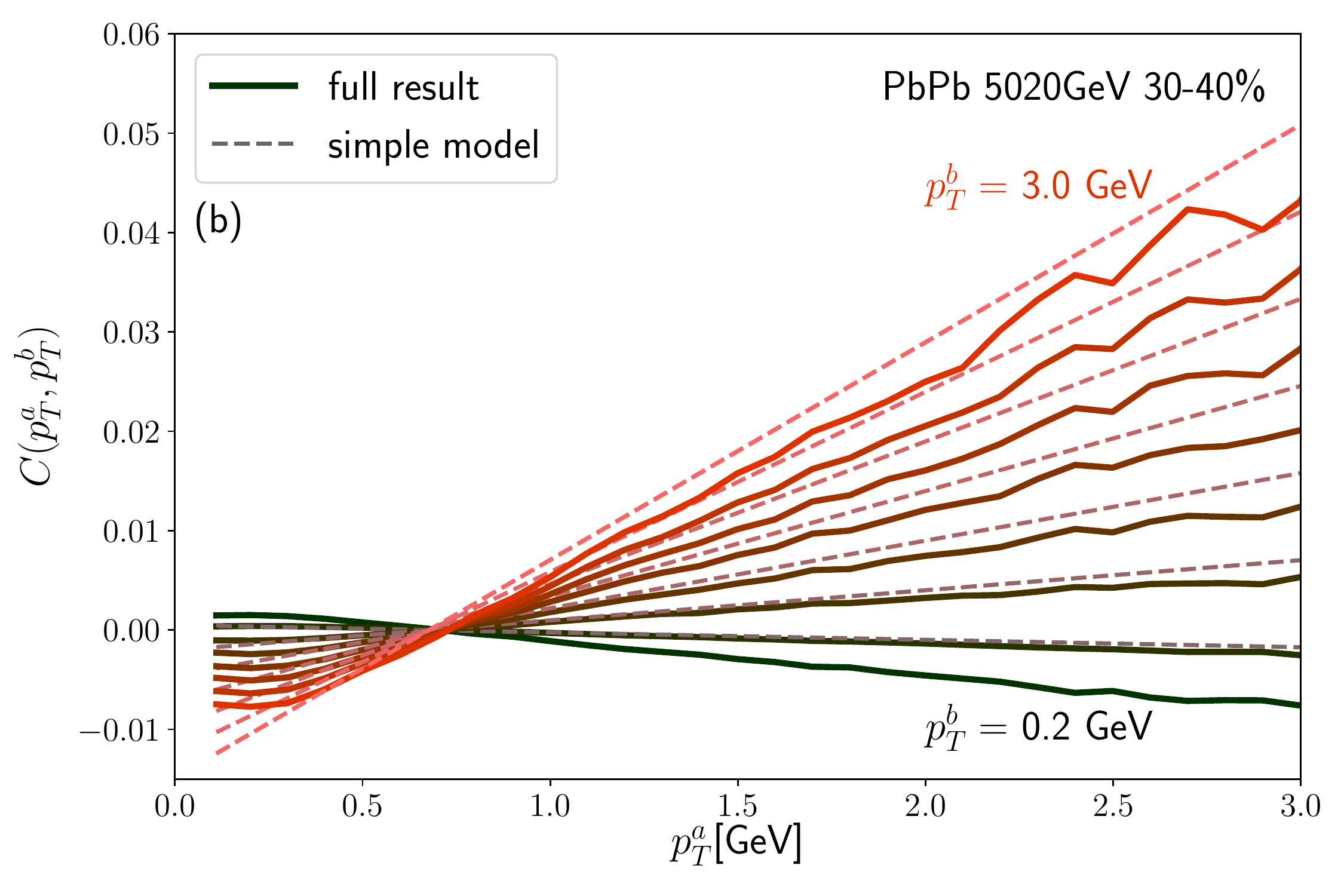}
\caption{ The correlation function $C(p_{T}^a,p_{T}^b)$ from \Eq{twoparticledef} from the IP-Glasma+\textsc{Music}+UrQMD calculation for Pb+Pb collisions at $\sqrt{s}=5020\,{\rm GeV}$ (solid lines) compared to (a) the factorized form $v_0(p_{T}^a)v_0(p_{T}^b)$ and (b) the result from the simple model of \Ref{Gardim:2019iah} given in \Eq{eq:modelresult} (dashed lines). Shown are 8 different $p_{T}^b$ values in steps of $0.4\,{\rm GeV}$. \label{fig:C-PbPb-C30-40-5020GeV-PID}}
\end{figure*}

The ALICE measure, $C_m/M(p_T)^2$, is closely related to  the $v_0$ measure.  
$v_0$ involves $\deltan P_T$ (as opposed 
to $\delta P_T$), which  subtracts the linear correlation with multiplicity, $ \llangle \delta P_T \delta N \rrangle /\sigma_N^2$.
With similar approximations to \Eq{Mofpt}, one finds that this correlation is
\begin{align}
   \frac{ \llangle \delta P_T \delta N \rrangle }{\sigma^2_N } \simeq& \frac{\llangle \delta \ptbar \delta N \rrangle }{\sigma^2_N } \ave{N} + \ave{\ptbar} \, ,
\end{align}
and thus 
\st 
\frac{\deltan P_T}{\llangle \ptbar \rrangle \llangle N \rrangle}  \simeq \frac{\deltan \ptbar}{\llangle\ptbar \rrangle} \, .
\stp
This differs from \Eq{eq:DPt2} by the ``hats''. Thus
\st
v_0^2 \simeq \frac{\sigman_{\ptbar}^2 }{\llangle \ptbar \rrangle^2} \, ,
\stp
and
\st
\label{Cmv0precise}
\frac{C_m}{M(p_T)^2 }   = v_0^2 + \frac{\llangle \delta \ptbar \delta N \rrangle^2 }{\llangle \ptbar \rrangle^2 \sigma_N^2} \, .
\stp
The last term in \Eq{Cmv0precise} is small if the multiplicity
bins are very narrow as was the case with the ALICE measurements.

Measuring both $C_m$ and $v_0^2$ allows access to the physical covariances, $\llangle \ptbar^2 \rrangle$ and  $\llangle \delta \ptbar \delta N \rrangle/\sigma_N^2 $.  In a principal component analysis of momentum fluctuations (described briefly in the next section), the combination of the leading and subleading principal components also gives access to these variances~\cite{Gardim:2019iah}.

\subsubsection{Principal components} \label{sec:principal}

\begin{figure*}[htb]
\includegraphics[width=0.48\textwidth]{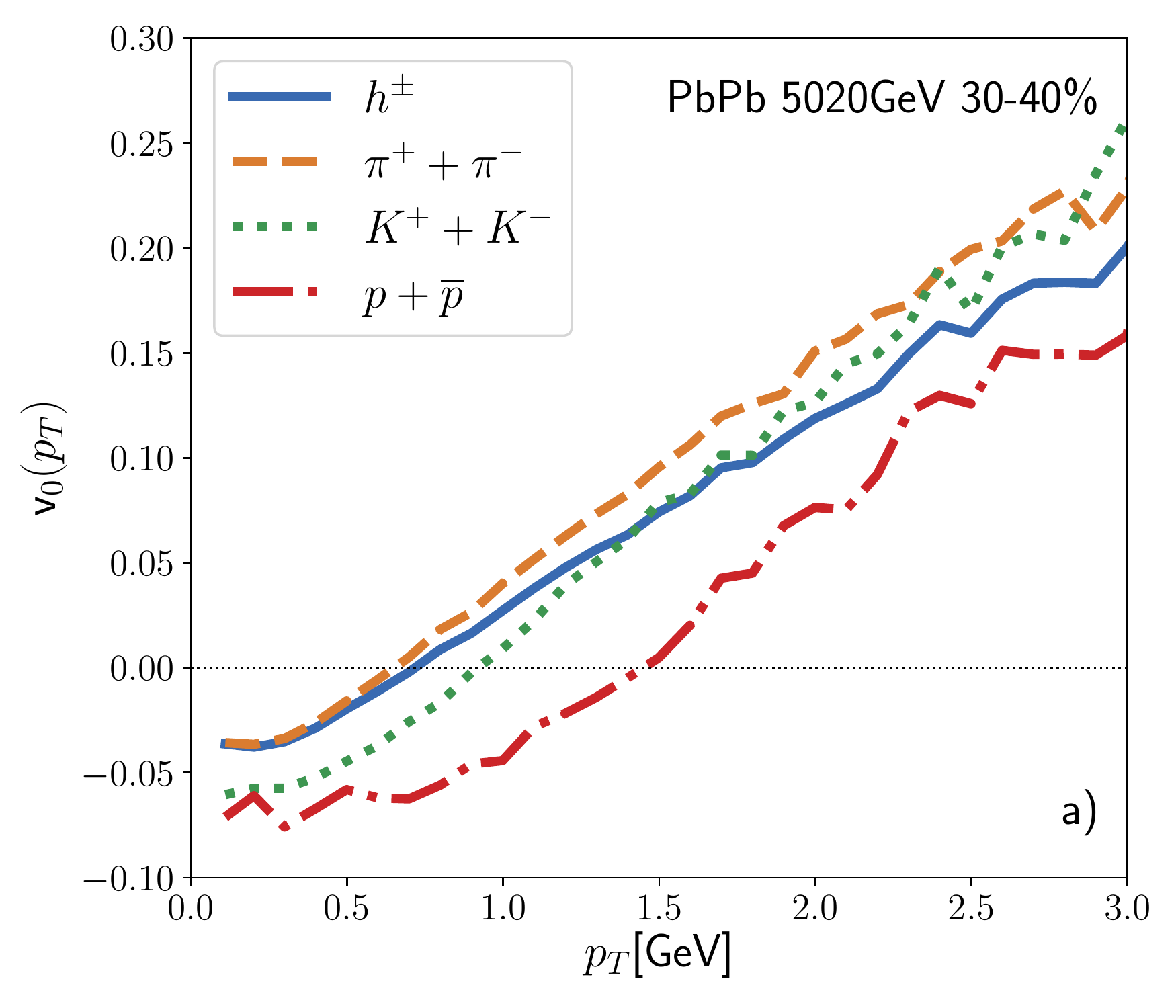} 
\includegraphics[width=0.48\textwidth]{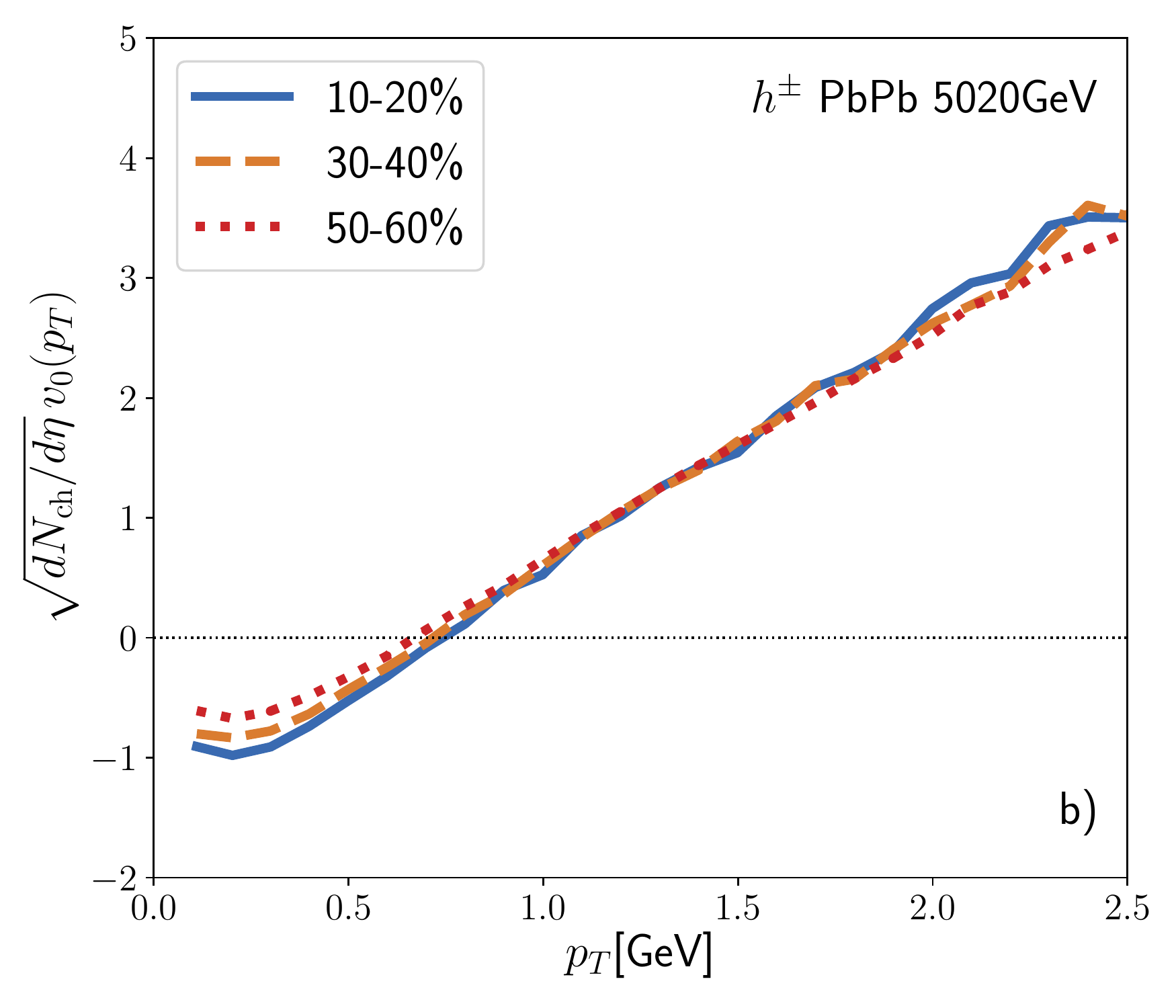}
  \caption{The momentum dependent $v_0(p_T)$ for charged hadrons (solid line) and identified particles (pions (dashed), kaons (dotted), protons (dash-dotted)) in 30-40\% central $\sqrt{s}=5020\,{\rm GeV}$ Pb+Pb collisions (a). Charged hadron $v_0(p_T)$ scaled by $\sqrt{dN_{\rm ch}/d\eta}$ in the respective centrality bin for three different centralities (b). \label{fig:v0-PbPb-C30-40-5020GeV-PID}}
\end{figure*}
Momentum fluctuations have been studied experimentally and theoretically using  principal components~\cite{Bhalerao:2014mua,Sirunyan:2017gyb}.
Briefly, the principal component  method breaks up the correlation matrix $\llangle \NN(p^a) \NN(p^b) \rrangle$  into eigen-vectors. The leading eigen-vector predominantly reflects multiplicity (or centrality) fluctuations $\llangle \delta N^2 \rrangle$, while the subleading eigen-vector predominantly reflects
momentum fluctuations $\llangle (\delta \ptbar)^2 \rrangle$~\cite{Mazeliauskas:2015efa}.
Further analysis explained that this description is only approximately true, and in general the leading and subleading eigen-vectors are mixtures of these two contributions~\cite{Gardim:2019iah}.  In terms of the model discussed in \Sect{sec:OlliModel} the leading and subleading modes are~\cite{Gardim:2019iah}
\begin{subequations}
\begin{align}
   v_0^{(1)}(p_T) =&  \frac{\sigma_N }{N} +  \left[  
      \frac{ 
         - \left(
      \frac{\sigma_{\ptbar}}{\llangle\ptbar \rrangle} \right)^2
      + 2 \frac{ \llangle \delta N \delta \ptbar \rrangle }{\llangle N \rrangle \llangle\ptbar \rrangle }  }
   { \frac{\sigma_N }{\llangle N \rrangle} } \right]  \frac{p_T}{\llangle \ptbar \rrangle} \, ,  \\
   v_0^{(2)}(p_T) =&   - \frac{3}{2} \frac{\sigma_\ptbar}{\llangle \ptbar \rrangle} \left(1 - \frac{4}{3} \frac{p_T}{\llangle \ptbar \rrangle } \right) \, ,
\end{align}
\end{subequations}
which should be compared to \Eq{eq:v0ptmodel}.
The $v_0^{(\alpha)}$ principal components have been measured by the CMS Collaboration~\cite{Sirunyan:2017gyb}, and  are
moderately well described by (ideal) hydrodynamic simulations~\cite{Gardim:2019iah}. We believe that the observable
$v_0(p_T)$ described here 
can help clarify the discrepancies with hydrodynamic simulations. Further, 
both theoretically and experimentally,
$v_0$ is a simpler measure
than the corresponding principal components, which often can mix different physics, such as flow and non-flow, in counter-intuitive ways.

\subsection{Simulations of momentum fluctuations}

\subsubsection{IP-Glasma+MUSIC+UrQMD}\label{sec:results}
In this section we present results for the observables discussed above using simulations of heavy ion collisions with the hybrid framework of IP-Glasma \cite{Schenke:2012wb,Schenke:2012hg}, \textsc{Music} \cite{Schenke:2010nt,Schenke:2010rr,Schenke:2011bn}, and UrQMD \cite{Bass:1998ca,Bleicher:1999xi}. IP-Glasma provides the initial energy momentum tensor computed from the gluon fields of the incoming nuclei that are described in the Color Glass Condensate framework \cite{McLerran:1994ni,McLerran:1994ka,Iancu:2003xm}. It includes fluctuations of nucleon positions and color charges, which manifest themselves in fluctuations of all components of the energy momentum tensor, leading e.g. to energy density and initial flow velocity fluctuations. \textsc{Music} describes the relativistic hydrodynamic evolution of the initial energy momentum tensor, including shear and bulk viscosity, with all parameters described in \cite{Schenke:2019ruo}. The low energy density regime ($e<0.18$ GeV/fm$^3$) is described microscopically as an interacting hadron resonance gas using UrQMD. Hadrons for this latter step of the simulation are obtained by first using the Cooper-Frye \cite{Cooper:1974mv} procedure to get the off-equilibrium momentum distributions \cite{Dusling:2009df,Bozek:2009dw,Paquet:2015lta} of all particle species and then sampling those distributions \cite{iSS}. For each hydrodynamic event we run many UrQMD events until the total number of particles per unit rapidity reaches at least 100,000, and use these to compute particle  spectra  and  flow vectors. This procedure ensures sufficient statistics and eliminates non-flow correlations from UrQMD.

\subsubsection{Results on transverse momentum fluctuations} 
\label{sec:pTfluct}
We begin by showing the correlator $C(p_{T}^a,p_{T}^b)$, defined in Eq.\,\eqref{twoparticledef}, from the hybrid IP-Glasma+\textsc{Music}+UrQMD calculation for Pb+Pb collisions at $\sqrt{s}=5020\,{\rm GeV}$ in Fig.\,\ref{fig:C-PbPb-C30-40-5020GeV-PID}. We show the $p_{T}^a$ dependence of the correlator for charged hadrons for 8 different values of $p_{T}^b$ using solid lines. To avoid self-correlations, we take the two parts of $C(p_{T}^a,p_{T}^b)$ in different rapidity intervals.

In Fig.\,\ref{fig:C-PbPb-C30-40-5020GeV-PID}a) dashed lines represent the factorized result $v_0(p_{T}^a)v_0(p_{T}^b)$, using $v_0(p_T)$ obtained from the correlator via
\begin{equation}
   v_0(p^a) = \frac{\int {\rm d}p^b \,  p^b  \langle \NN(p^b)\rangle \,   C(p^a,p^b)}{\sqrt{\int {\rm d}p^a {\rm d} p^b  \, p^a p^b \langle \NN(p^a) \rangle \langle \NN(p^b)\rangle   C(p^a,p^b)}}, \notag 
\end{equation}
where integrations run from $p_T^{a/b}=0$ to a maximum of $p_T^{a/b}=4\,{\rm GeV}$.
The result demonstrates that indeed the result factorizes to a good approximation, with significant deviations appearing only for the largest momenta $p_T^a\sim p_T^b \sim 3\,{\rm GeV}$. Similar results were found for other centrality classes. 

In Fig.\,\ref{fig:C-PbPb-C30-40-5020GeV-PID}b) we again show the calculation's result for the correlator $C(p_T^a,p_T^b)$ as solid lines, this time compared to the simple model Eq.\,\eqref{eq:modelresult}. Agreement between the simple model, which produces linear behavior with $p_T^a$ and $p_T^b$ is also rather good, except at the lowest momenta. Also, the model slightly overestimates the full result for most $p_T^b$ when $p_T^a\gtrsim 1\,{\rm GeV}$. We expect all lines to cross at $p_T^a=\langle \bar{p}_T \rangle$, which is an exact result in the model Eq.\,\eqref{eq:modelresult}, and we find the full result to be very close to that.

We show $v_0(p_T)$ from the hybrid IP-Glasma+\textsc{Music}+UrQMD calculation for 30-40\% central Pb+Pb collisions at $\sqrt{s}=5020\,{\rm GeV}$ in Fig.\,\ref{fig:v0-PbPb-C30-40-5020GeV-PID}a). Besides the results for charged hadrons (solid lines) extracted from the correlator $C(p_{T}^a,p_{T}^b)$ discussed above, we also present $v_{0,s}(p_T)$ for identified particles (pions, kaons, protons), defined in \Eq{eq:v0speciesdef}.
A clear species dependence is visible, with the result for $v_0(p_T)$ crossing zero at approximately the mean transverse momentum of the respective particle. The charged hadron $v_0(p_T)$ is a weighted average of the individual charged particle $v_{0,s}(p_T)$.

To show the centrality dependence of the signal, and how it scales with $dN_{\rm ch}/d\eta$, in Fig.\ref{fig:v0-PbPb-C30-40-5020GeV-PID}b) we show the charged hadron $v_0(p_T)$ for three different centralities, multiplied by the $\sqrt{dN_{\rm ch}/d\eta}$ in the respective centrality bin. We see that most of the centrality dependence of $v_0(p_T)$ is a result of the fluctuations changing as $(\sqrt{dN_{\rm ch}/d\eta})^{-1}$, with small deviations from this scaling at $p_T<0.5\,{\rm GeV}$ caused by the difference in averaged radial flow in different centrality bins.

To determine how strongly the transverse momentum fluctuations correlate with
the initial geometry, we introduce predictors for $\hat{\delta}P_T$, motivated
by earlier work~\cite{Broniowski:2009fm,Bozek:2012fw,Bozek:2017elk}. First, we 
correlated the fluctuations in the  total transverse momentum $\deltan P_T$ 
with the average entropy density in a given event $[s]$, with the average computed as 
\st
\label{barsdefined}
[f] = \frac{\int d\xt e(\xt) f(\xt)}{\int d\xt e(\xt) } \, .
\stp
Here $e(\xt)$ is
the initial energy density and $f(\xt)$ is the quantity to be averaged over the transverse plane. Note that we recenter such that $[\xt]=(0,0)$.

Next, we correlated $\deltan P_T$ with the fluctuations in total entropy per area, where the circular area is given by $\pi [r^2]$. Including the elliptic deformation in the definition of area can improve this predictor significantly \cite{Mazeliauskas:2015efa,Bozek:2017elk}. Thus we define the 
elliptic area as 
\begin{equation}
A \equiv \pi [r^2] \sqrt{1 - \varepsilon_2^2} = \pi \sqrt{4[x^2][y^2]}  \, ,
\end{equation}
where $\varepsilon_2 \equiv ([y^2] - [x^2])/([x^2]+[y^2])$,  and $x$ and $y$  are measured along the short and long principal axes of the event-by-event ellipse.

To quantify the quality of the predictors, we define the Pearson correlation coefficient 
\begin{equation}
    Q_\xi = \frac{\langle \hat{\delta}P_T \hat{\delta}\xi\rangle}{\sqrt{\langle \hat{\delta}P_T^2 \rangle \langle \hat{\delta}\xi^2 \rangle}}\,,
\end{equation}
where the predictor $\hat{\delta}\xi$ can be defined in various different ways as discussed above. We will use the average event entropy density $\xi = [s]$,  the entropy per circular area  $\xi=S/(\pi[r^2])$, and the entropy per   elliptic area $\xi=S/A$.
Finally, we use $\xi=[r^2]$ as suggested in the original work \cite{Broniowski:2009fm}, which naturally produces a negative correlation, and is expected to be the least efficient estimator.

\begin{figure}[htb]
\includegraphics[width=0.48\textwidth]{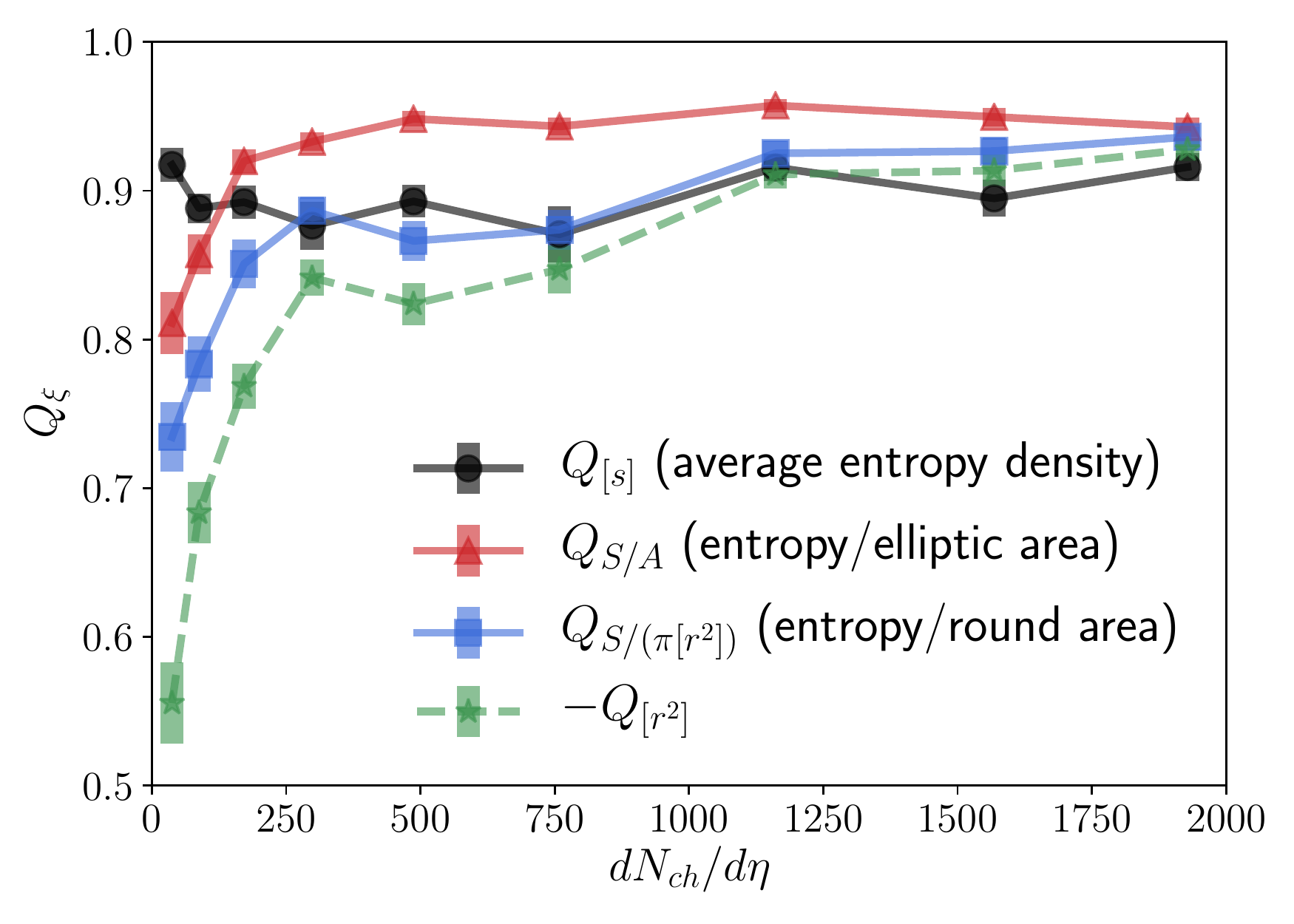}
\caption{The Pearson coefficients measuring the quality of different initial state estimators for transverse momentum fluctuations in $\sqrt{s}=5020\,{\rm GeV}$ Pb+Pb collisions.  \label{fig:Q-differentEstimators-PbPb5020}}
\end{figure}

Results for the different Pearson coefficients $Q_\xi$ are shown in
Fig.\,\ref{fig:Q-differentEstimators-PbPb5020}. We find that for the most
central collisions, all predictors work approximately equally well. With
decreasing multiplicity, the interaction region becomes more elliptic and
taking into account the eccentricity of the area improves the predictor using
entropy per area dramatically. 
The entropy per elliptic area is a simple one term predictor, which 
   combines the most important physics and  works well over the full range in centrality. 
   It should be compared with the multi-term predictors of Refs.~\cite{Mazeliauskas:2015efa,Bozek:2020drh},  which separately included the fluctuations in radius, entropy, and squared eccentricities to achieve a similar correlation coefficient. 

   In rather peripheral collisions, when the eccentricity is large $\varepsilon_2 \sim 1$  
the interaction region typically consists of several disconnected regions and 
not one single (approximate) ellipse.   To increase the mean $p_T$ in this regime it is better 
to have several clustered hot spots leading to a strong hydrodynamic response.
Thus, the average entropy density $[s]$ provides a better predictor $p_T$ in peripheral bins.

Motivated by \Ref{Gardim:2020sma}, we investigated their preliminary proposal to use the total initial energy (at fixed multiplicity) as a predictor and found it comparable to $[s]$, except in very peripheral events where using $[s]$ works better. We also investigated $E/A$ and found it marginally better than $S/A$.

\subsection{Comparison with existing data} 
\label{Sect:ComparisonWpt} 

\begin{figure}[tb]
\includegraphics[width=0.48\textwidth]{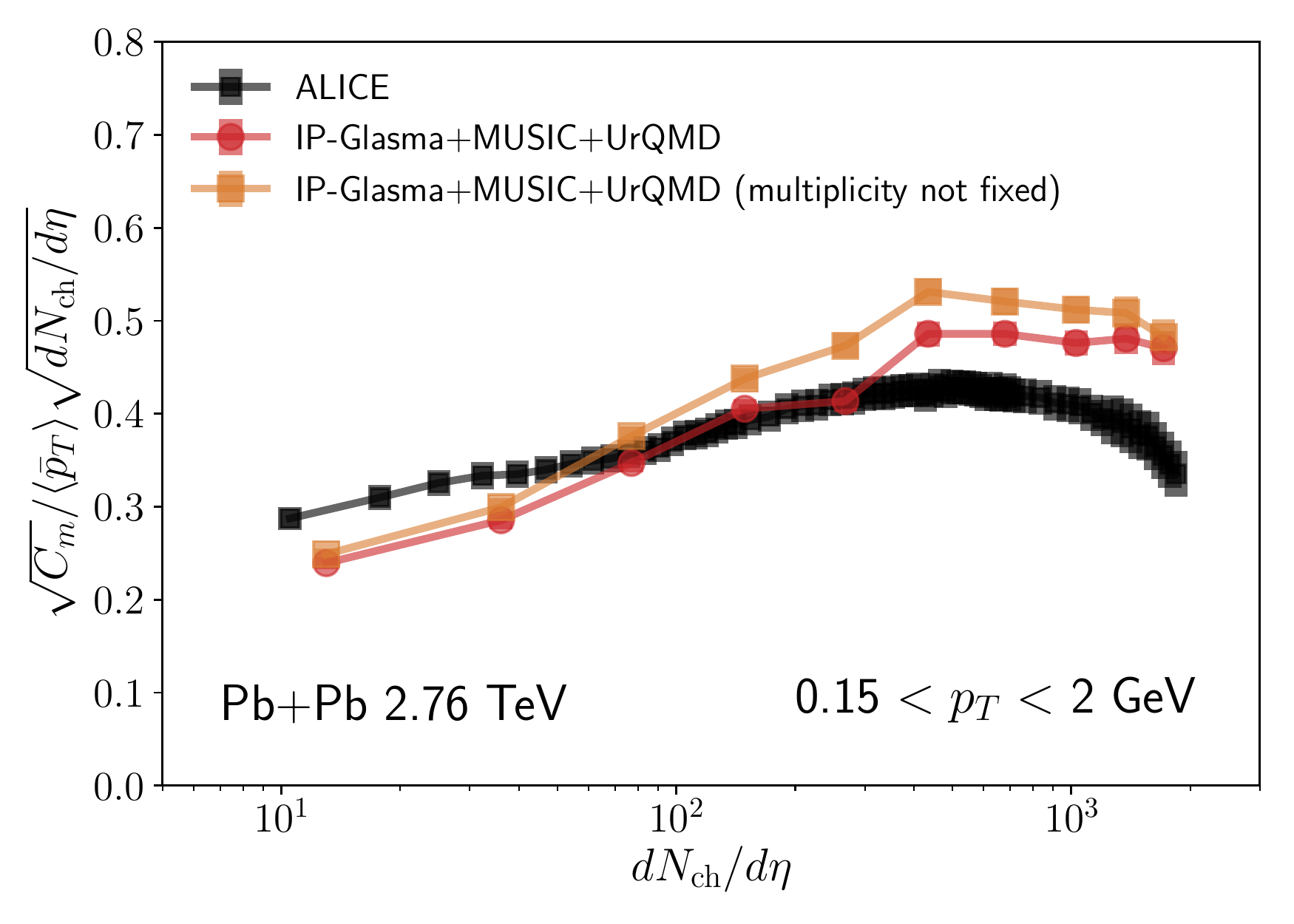}
  \caption{The observable $\sqrt{C_m}/\langle\bar{p}_T\rangle$ scaled by $\sqrt{dN_{\rm ch}/d\eta}$ in Pb+Pb collisions at $\sqrt{s}=2760\,{\rm GeV}$ compared to experimental data from the ALICE Collaboration \cite{Abelev:2014ckr}. Results are computed with (circles) and without (squares) fixing the multiplicity. The simulation result for $\sqrt{C_m}/\langle\bar{p}_T\rangle$ at fixed multiplicity (circles) should be compared to the data, and is  approximately equal to $v_0 \sqrt{dN_{\rm ch}/d\eta}$. See \Eq{Cmv0precise} for the precise relationship between $v_0$ and $C_m$. \label{fig:CmOverMpT-PbPb-2760GeV}}
\end{figure}

We move on to compare our results for transverse momentum fluctuations to
existing experimental measurements.  In Fig.\,\ref{fig:CmOverMpT-PbPb-2760GeV}
we show the observable $\sqrt{C_m}/\langle\bar{p}_T\rangle$ scaled by
$\sqrt{dN_{\rm ch}/d\eta}$ in Pb+Pb collisions at $\sqrt{s}=2760\,{\rm GeV}$
and compare to the result from the ALICE Collaboration \cite{Abelev:2014ckr}.
The $p_T$ range used to compute the observable is $0.15 < p_T < 2\,{\rm GeV}$,
while the multiplicity is computed over the entire $p_T$ range. We show both
the regular $\sqrt{C_m}/\langle\bar{p}_T\rangle$, which we compute using the
approximation in Eq.\,\eqref{eq:CmApprox}, and the same quantity but with the
multiplicity fixed. The latter should compare better to the experimental
result, as the multiplicity bins used by the ALICE Collaboration are very
narrow. Fixing the multiplicity decreases the fluctuations by removing those
resulting from multiplicity fluctuations, and improves agreement with the
experimental data. 
The slope of our result as a function of the multiplicity $dN_{\rm ch}/d\eta$ is steeper than the experimental data, a trend we will also see in the observable $c_k$ measured by the ATLAS Collaboration \cite{Aad:2019fgl} (see \Fig{fig:ATLAS-ck-PbPb5020}).

\begin{figure}[htb]
\includegraphics[width=0.48\textwidth]{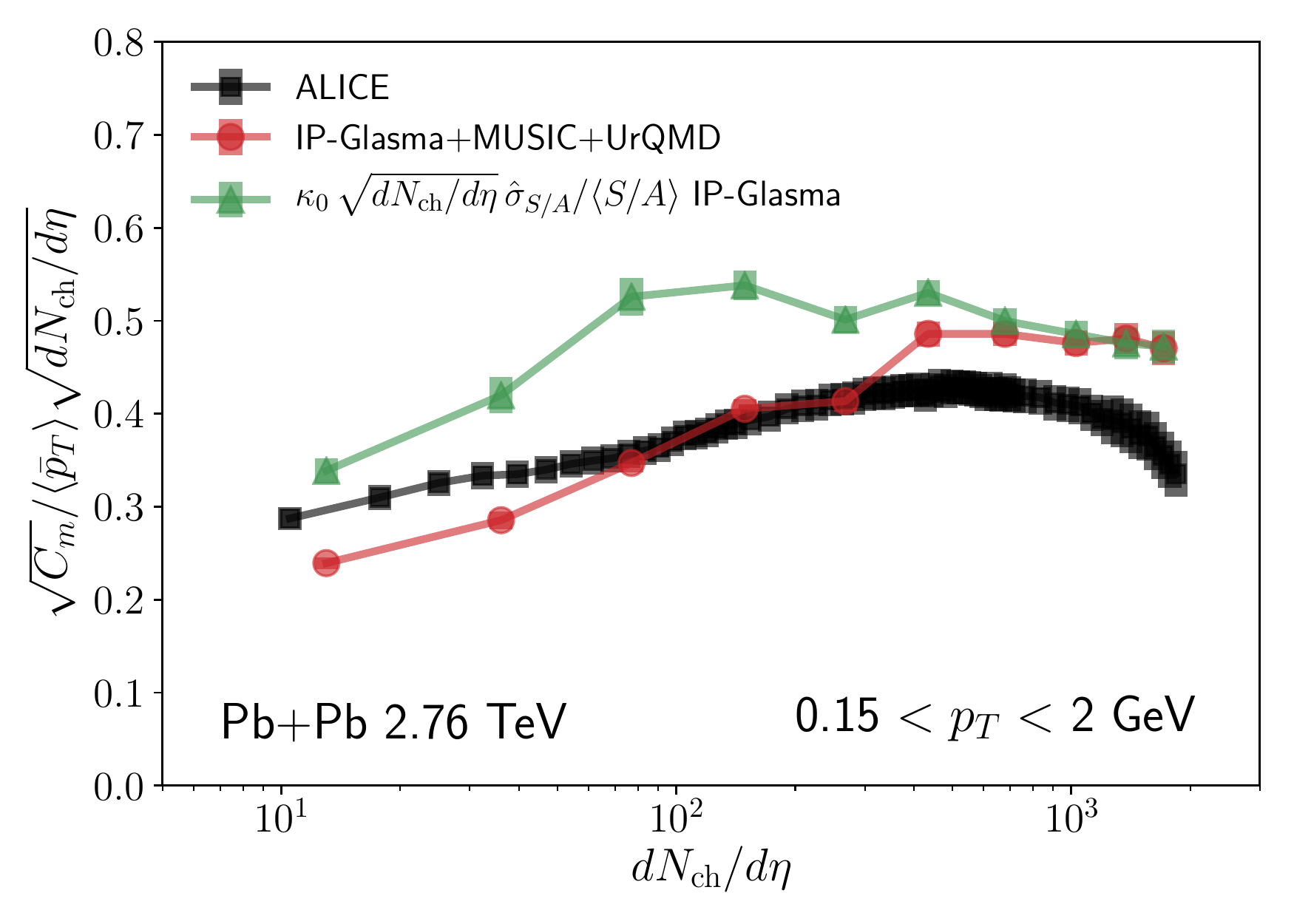}
  \caption{The observable $\sqrt{C_m}/\langle\bar{p}_T\rangle$ at fixed multiplicity scaled by $\sqrt{dN_{\rm ch}/d\eta}$ in Pb+Pb collisions at $\sqrt{s}=2760\,{\rm GeV}$ (circles) compared to experimental data from the ALICE Collaboration \cite{Abelev:2014ckr}. Also shown
  are the initial state fluctuations $\kappa_0 \hat{\sigma}(S/A)/\langle S/A \rangle$ (triangles) taken from the IP-Glasma initial conditions (See \Eq{eq:naivekappa0}). \label{fig:CmOverMpT-PbPb-2760GeV-Estimator}}
\end{figure}

The multiplicity dependence of
the model
follows both from the centrality dependence of the initial state fluctuations of  $S/A$, and the hydrodynamic response coefficient $\kappa$
\st
\frac{\deltan P_T}{\llangle P_T \rrangle}  = \kappa  \frac{\deltan (S/A) }{\llangle S/A \rrangle } \, .
\stp
We would like to disentangle these two contributions.
In central collisions the response coefficient
is approximately $\kappa_0 \simeq 0.133$. 
If $\kappa$ were constant with centrality,  the model's $P_T$ fluctuations would follow the initial conditions alone
\st
\label{eq:naivekappa0}
v_0 = \kappa_0 \,  \frac{ \hat \sigma_{S/A} }{\llangle S/A \rrangle} \, .
\stp
Fig.\,\ref{fig:CmOverMpT-PbPb-2760GeV-Estimator} shows this somewhat naive prediction of the IP-Glasma model. 
Interestingly, the initial state estimator given in \Eq{eq:naivekappa0} reproduces the shape of the experimental measurement somewhat better than the full result of our hybrid framework calculation (also shown in Fig.\,\ref{fig:CmOverMpT-PbPb-2760GeV-Estimator}). This might indicate that details of the final state evolution, such as temperature dependent transport coefficients and the freezeout prescription, have not been modeled optimally.

\begin{figure}[tb]
\includegraphics[width=0.48\textwidth]{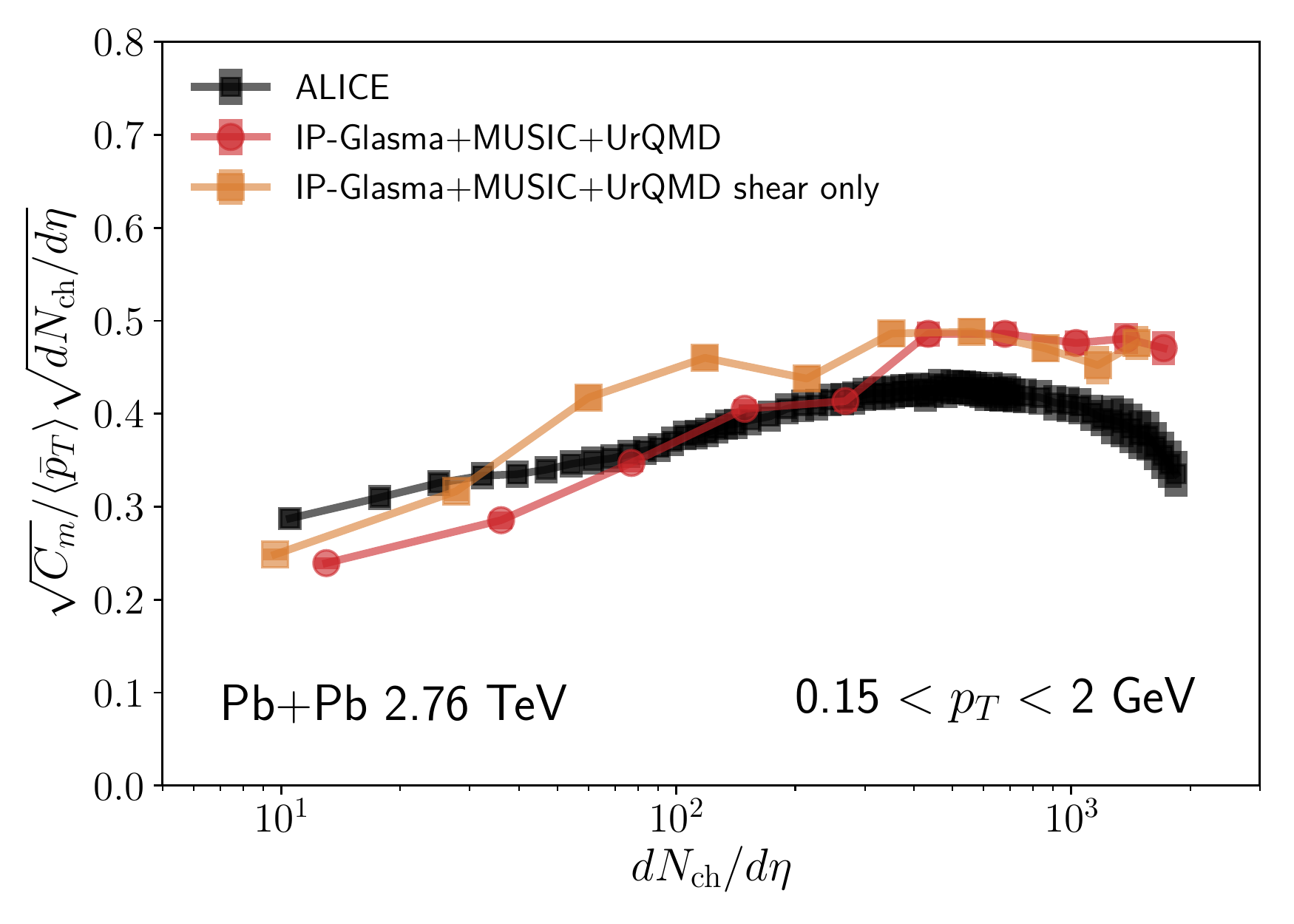}
  \caption{The observable $\sqrt{C_m}/\langle\bar{p}_T\rangle$ at fixed multiplicity scaled by $\sqrt{dN_{\rm ch}/d\eta}$ in Pb+Pb collisions at $\sqrt{s}=2760\,{\rm GeV}$ (circles) compared to experimental data from the ALICE Collaboration \cite{Abelev:2014ckr}, and the result from using no bulk viscosity in the calculation (squares).\label{fig:CmOverMpT-PbPb-2760GeV-shearonly}}
\end{figure}

\begin{figure}[htb]
\includegraphics[width=0.48\textwidth]{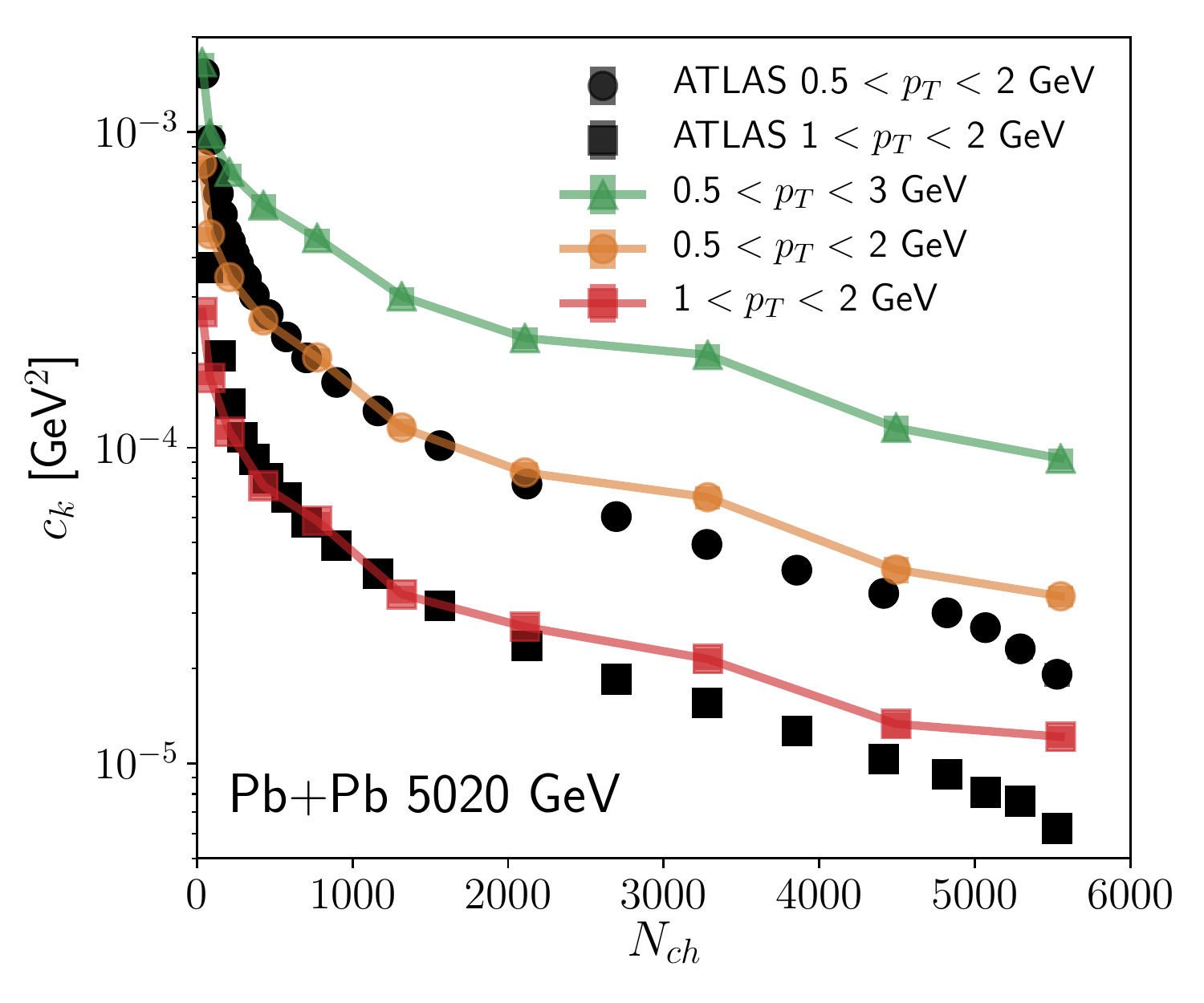}
  \caption{The observable $c_k$, computed here as $c_k = \hat{\sigma}_{\bar{p}_T}^2$ compared to experimental data from the ATLAS Collaboration \cite{Aad:2019fgl}. We show results for three different $p_T$ ranges, two of which were also measured by ATLAS. \label{fig:ATLAS-ck-PbPb5020}}
\end{figure}

To verify whether that is indeed a possibility, we compute 
$\sqrt{C_m}/\langle\bar{p}_T\rangle$ using a simulation with shear viscosity
only. As shown in Fig.\,\ref{fig:CmOverMpT-PbPb-2760GeV-shearonly}, in this
case the multiplicity dependence of $\sqrt{C_m}/\langle\bar{p}_T\rangle$ is
indeed different from the full result including bulk viscosity, which shows
that mean-$p_T$ fluctuations should be considered when constraining transport
coefficients in the future. The shear-only result overestimates the fluctuation data in peripheral collisions, and also
yields a $\langle\bar{p}_T\rangle$ that is too large~\cite{Ryu:2015vwa}. Nevertheless, this exercise demonstrates the sensitivity of mean transverse momentum fluctuations to the choice of
transport coefficients, and thus further highlights their importance. 

Next, in Fig.\,\ref{fig:ATLAS-ck-PbPb5020} we compare to the observable $c_k$ as a function of multiplicity, which when measured in narrow multiplicity bins, should be well approximated by $\hat{\sigma}_{\bar{p}_T}^2$. To be able to compare to the ATLAS measurement \cite{Aad:2019fgl}, we compute  $\hat{\sigma}_{\bar{p}_T}^2$ with $\bar{p}_T$ and $\bar{p}_T^2$ determined in the ranges $0.5 < p_T < 2\,{\rm GeV}$ and $1 < p_T < 2\,{\rm GeV}$, with the multiplicity fixed in the interval $0.5 < p_T < 5\,{\rm GeV}$. In ATLAS, $N_{\rm ch}$ is determined over 5 units of rapidity, such that we use $N_{\rm ch}= 5 \, \int dp\, \NN(p)$. Note that we chose to determine $c_k$ at fixed multiplicity for this comparison because the multiplicity bins used by ATLAS are significantly narrower than the ones we used. 

Agreement with the ATLAS result is reasonable in the range $200<N_{\rm ch}<2000$, with the experimentally measured $c_k$ being underestimated at smaller $N_{\rm ch}$ and overestimated at larger values. This is the same behavior as we observed for $\sqrt{C_m}/\langle\bar{p}_T\rangle$ above. The $p_T$-cut dependence is also qualitatively reproduced. We note that using different $p_T$ ranges for this observable leads to changing $\langle \bar{p}_T \rangle$ as well as changing $p_T$ fluctuations. It would be much more straight forward to simply measure $v_0(p_T)$ and the factorization of the $C(p_T^a,p_T^b)$ to gain insight into the $p_T$ dependence of fluctuations in the particle spectrum. 

The IP-Glasma{+}\textsc{Music}{+}UrQMD hybrid framework switches abruptly from the off-equilibrium Yang-Mills system of the IP-Glasma model to viscous hydrodynamics. A more realistic transition between the two stages of the evolution can be obtained by including a pre-equilibrium stage provided by the KoMPoST framework \cite{Kurkela:2018vqr,Kurkela:2018wud}. This framework uses non-equilibrium Green’s functions from QCD kinetic theory to propagate the IP-Glasma energy-momentum tensor to the hydrodynamic stage, smoothly extrapolating the late stages of the IP-Glasma setup to hydrodynamics where the \textsc{Music} code takes over. Including the pre-equilibrium stage, the complete model consists of IP-Glasma{+}KoMPoST{+}\textsc{Music}{+}UrQMD.

When including a KoMPoST stage, connecting IP-Glasma and \textsc{Music} over the interval\footnote{In the notation of \cite{Kurkela:2018wud} $\tau_{\rm EKT}=0.1\,{\rm fm}$ and $\tau_{\rm hydro}=0.8\,{\rm fm}$. See Fig.~3 of that reference for an overview.}, $\tau = 0.1 - 0.8\,$fm/$c$, we need to modify the used temperature dependent bulk viscosity over entropy density ratio $(\zeta/s)(T)$, to improve the agreement of $\langle \bar{p}_T \rangle$ with experimental data. The pre-equilibrium dynamics in KoMPoST leads to more radial flow, hence the peak value of $(\zeta/s)(T)$ was increased by 35\% \cite{Gale:2020xlg}.

\begin{figure}[htb]
\includegraphics[width=0.48\textwidth]{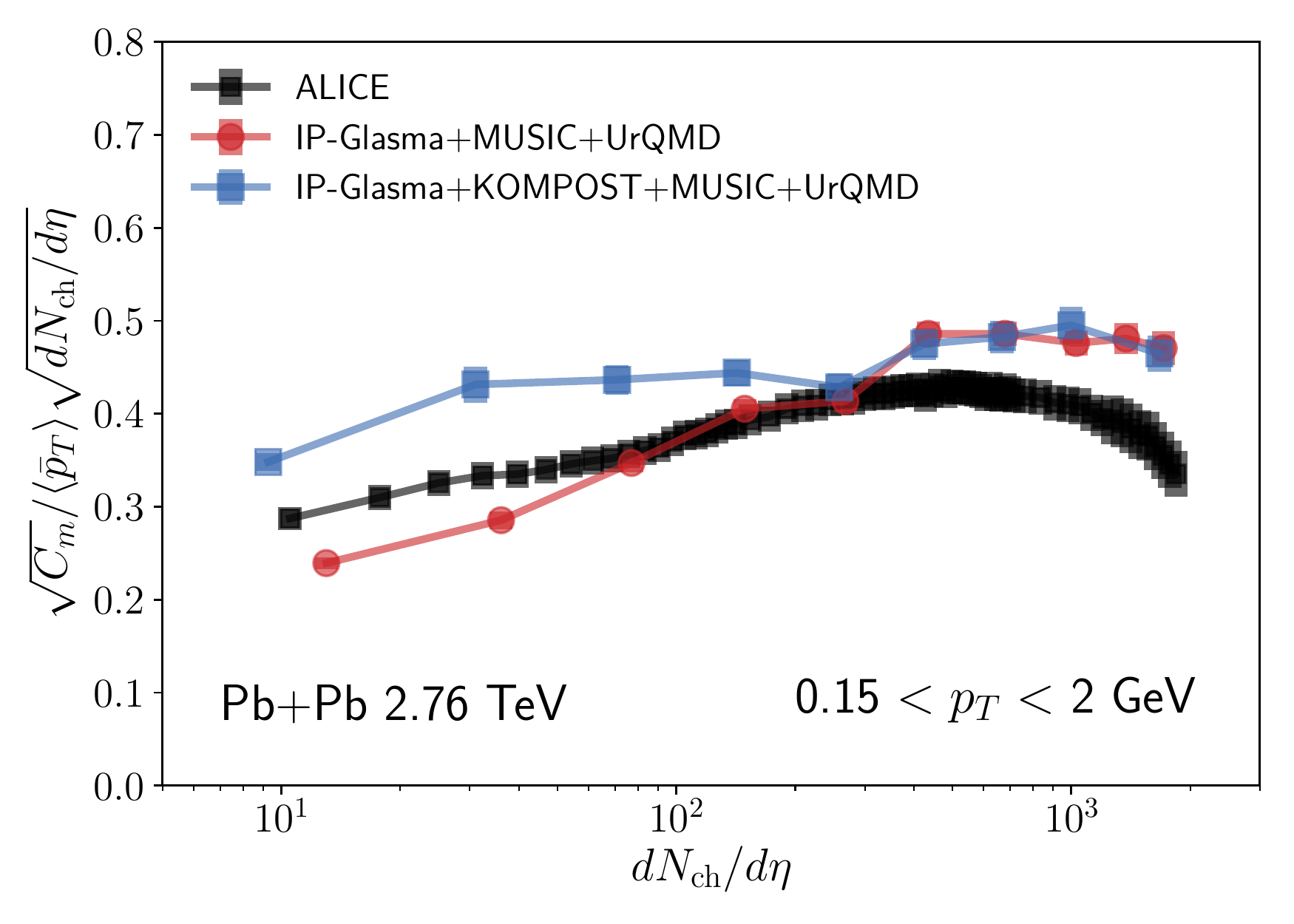}
  \caption{The observable $\sqrt{C_m}/\langle\bar{p}_T\rangle$ at fixed multiplicity scaled by $\sqrt{dN_{\rm ch}/d\eta}$ in Pb+Pb collisions at $\sqrt{s}=2760\,{\rm GeV}$ (circles) compared to experimental data from the ALICE Collaboration \cite{Abelev:2014ckr}, and the result obtained from the simulation including the KoMPoST pre-equilibrium stage (squares).\label{fig:CmOverMpT-PbPb-2760GeV-KOMPOST}}
\end{figure}

We present the effect of including a KoMPoST stage in our hybrid model on the measure of mean transverse momentum fluctuations, $\sqrt{C_m}/\langle\bar{p}_T\rangle$, at fixed multiplicities, in Fig.\,\ref{fig:CmOverMpT-PbPb-2760GeV-KOMPOST}. We find that mean transverse momentum fluctuations, while largely unchanged for the 6 largest multiplicity bins (0-50\% centrality), increase at low multiplicities when using KoMPoST.

\section{Correlations of anisotropic flow and mean transverse momentum} \label{sec:v2pTcorrelations}

In this section, we extend our studies to the correlation between the event-by-event deviation of the squared elliptic flow coefficient and the event-by-event deviation of the mean $p_T$, both at fixed multiplicity, using the correlator
\begin{equation}\label{eq:v2PT}
    \hat{\rho}(v_2^2,\bar{p}_T)=\frac{\langle \deltan v_2^2 \deltan \bar{p}_T\rangle}{\sqrt{\langle(\deltan v_2^2)^2\rangle\langle(\deltan \bar{p}_T)^2\rangle}}\,.
\end{equation}

The correlation between $v_2$ and $p_T$ was first studied with principal components in \cite{Mazeliauskas:2015efa}, with the prediction that these correlations would drop dramatically in central collisions.
The specific correlator in \Eq{eq:v2PT} was developed in \Ref{Bozek:2016yoj} and simplifies the earlier proposal. 
Further, it was found that $\hat \rho(v_2^2,\bar{p}_T)$ is sensitive to the compactness of the source in proton-nucleus collisions, and thus can be used to differentiate initial state models~\cite{Bozek:2016yoj,Bozek:2020drh}.

To study how well initial state properties determine the final observable, we
construct a predictor for $\hat{\rho}(v_2^2,\bar{p}_T)$. We 
estimate
the fluctuations of the transverse momentum at fixed multiplicity 
with $\hat{\delta}(S/A)$, which showed the strongest correlation for most centralities, as
shown in Fig.\,\ref{fig:Q-differentEstimators-PbPb5020}. 
   We emphasize that since the total initial entropy  is tightly correlated with multiplicity (which is held fixed),  $\hat \delta(S/A)$ is nearly equivalent to $\hat \delta (1/A)$, where $A\equiv \pi [r^2] \sqrt{1- \varepsilon_2^2}$ is the elliptic area. 
We further use
$\hat{\delta}\varepsilon_2$ to estimate the elliptic flow fluctuations, and define the
predictor
\begin{equation}\label{eq:v2PTest}
    \hat{\rho}_{\rm est}(v_2^2,\bar{p}_T)=\frac{\langle \deltan \varepsilon_2^2 \deltan(S/A)\rangle}{\sqrt{\langle(\deltan \varepsilon_2^2)^2\rangle\langle(\deltan(S/A))^2\rangle}}\,,
\end{equation}
which should track 
$\hat\rho(v_2^2,\bar p_T)$.

\subsection{Results from IP-Glasma+MUSIC+UrQMD}
\begin{figure}[tb]
\includegraphics[width=0.48\textwidth]{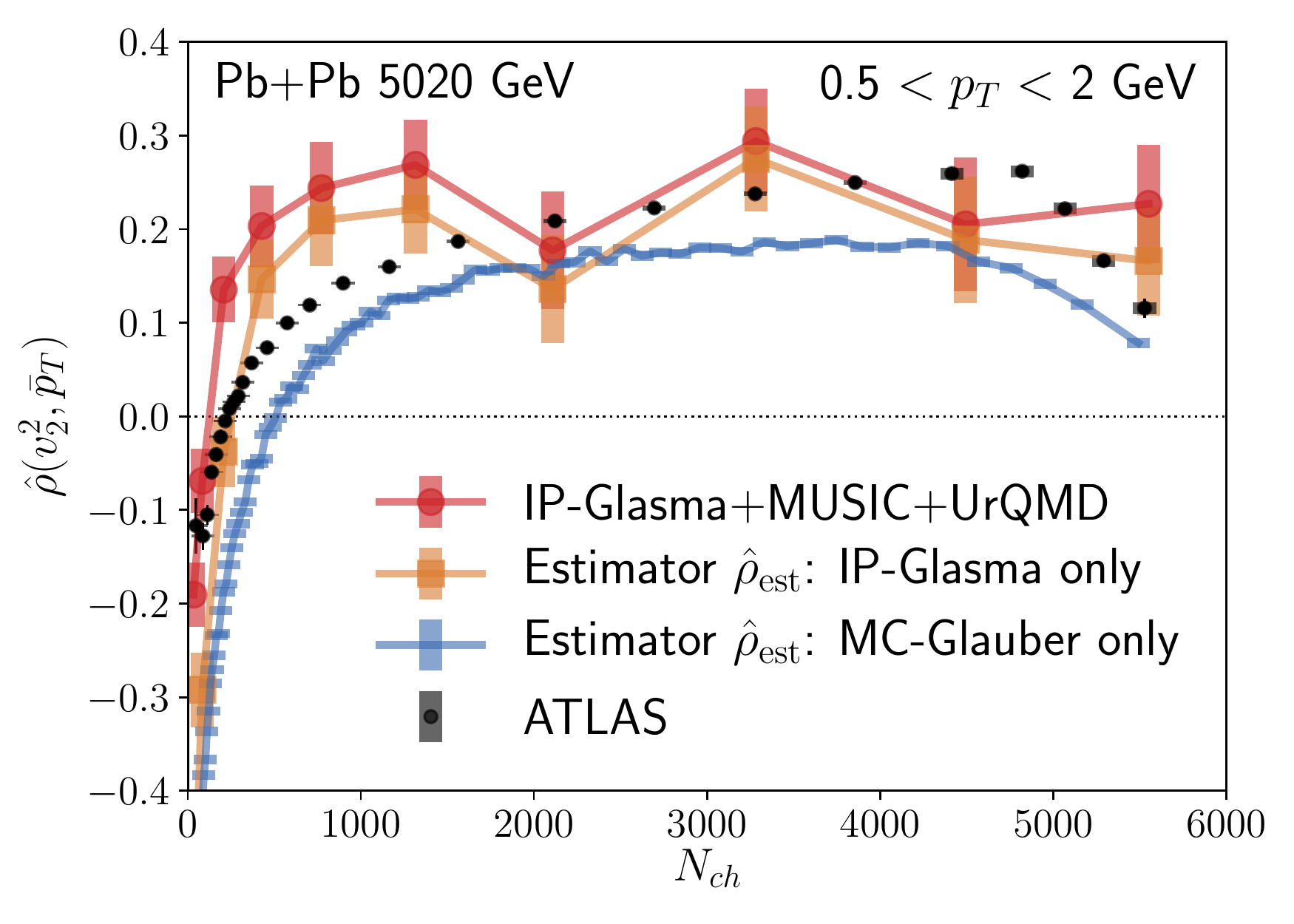}
\caption{The correlation measure between $v_2^2$ and mean transverse momentum $\hat{\rho}(v_2^2,\bar{p}_T)$ as a function of multiplicity compared to experimental data from the ATLAS Collaboration \cite{Aad:2019fgl} in $\sqrt{s}=5020\,{\rm GeV}$ Pb+Pb collisions. We further show $\hat{\rho}_{\rm est}$ from the IP-Glasma (squares) and MC-Glauber (lines) initial state models, using the eccentricity and entropy per elliptic area as predictors for $v_2$ and mean $p_T$, respectively.   \label{fig:ATLAS-rho-corr-PbPb5020}  
}
\end{figure}

In Fig.\,\ref{fig:ATLAS-rho-corr-PbPb5020} we present results from the IP-Glasma+\textsc{Music}+UrQMD hybrid model calculation for $\hat{\rho}(v_2^2,\bar{p}_T)$ in $\sqrt{s}=5020\,{\rm GeV}$ Pb+Pb collisions, along with the predictor $\hat{\rho}_{\rm est}$ as a function of the ATLAS $N_{\rm ch}$, and compare to ATLAS experimental data \cite{Aad:2019fgl}.

The expected drop in central collisions
is seen in the experimental data, but could not be resolved in the model with
current statistics.  However,  the sign change of the
correlator at low multiplicity is qualitatively reproduced. Because
the estimator also reproduces this behavior, we conclude that the sign
change is a geometrical effect reflecting the correlation between the inverse area $1/A$ and $\varepsilon_2$ at fixed multiplicity. 
Indeed, in mid-central collisions a smaller 
area at fixed multiplicity
is achieved  by fluctuating to larger impact parameters,
increasing $\varepsilon_2$.  The smaller area and correspondingly larger
$\varepsilon_2$  in this case yields
a positive correlation between $\ptbar$ and $v_2$. By contrast, in quite peripheral collisions a
smaller area at fixed multiplicity is achieved by clustering the participants
in a single region. This region will be less elliptic through the
clustering process.  %Thus, 
In the clustered case then, 
the smaller area and correspondingly smaller $\varepsilon_2$ yield
 a negative correlation between $\ptbar$ and $v_2$.

Because the initial state predictor can approximate the full results very well, it is worth comparing to the predictor obtained from a Monte-Carlo (MC) Glauber model with high statistics. To that effect we generate 4 million minimum bias Pb+Pb events at 5020 GeV using an open-source code package \textsc{superMC}\footnote{We use the current version of the \textsc{superMC} code in the github repository, \url{https://github.com/chunshen1987/superMC}.}, which is a part of the iEBE-VISHNU framework \cite{Shen:2014vra}. This MC-Glauber model assumed that the system's entropy density is proportional to a mixture of wounded nucleon and binary collision profiles. Multiplicity fluctuations in the local entropy density of individual wounded nucleons and binary collisions were introduced according to a Gamma distribution, which was fitted to the normalized multiplicity distribution in p+p collisions assuming Koba-Nielsen-Olesen (KNO) scaling \cite{Koba:1972ng}. We use the lattice equation of state from the hotQCD Collaboration \cite{Bazavov:2014pvz} to convert the initial entropy density to energy density, and compute the elliptic area $A$ and eccentricity $\varepsilon_2$. All the detailed parameters are listed in Ref.~\cite{Shen:2014vra}. With high statistics, we find that the model estimator decreases in the most central collisions because the eccentricity decreases faster compared to changes in $A$. This estimator reproduces the shape of the $\hat{\rho}$ correlation function very well, although it underestimates the absolute strength by $\sim20\%$. The difference between the estimators from MC-Glauber and IP-Glasma initial conditions reflects that this observable is sensitive to the details of the initial condition, such as the degree of sub-nucleon fluctuations.

\begin{figure}[h]
\includegraphics[width=0.48\textwidth]{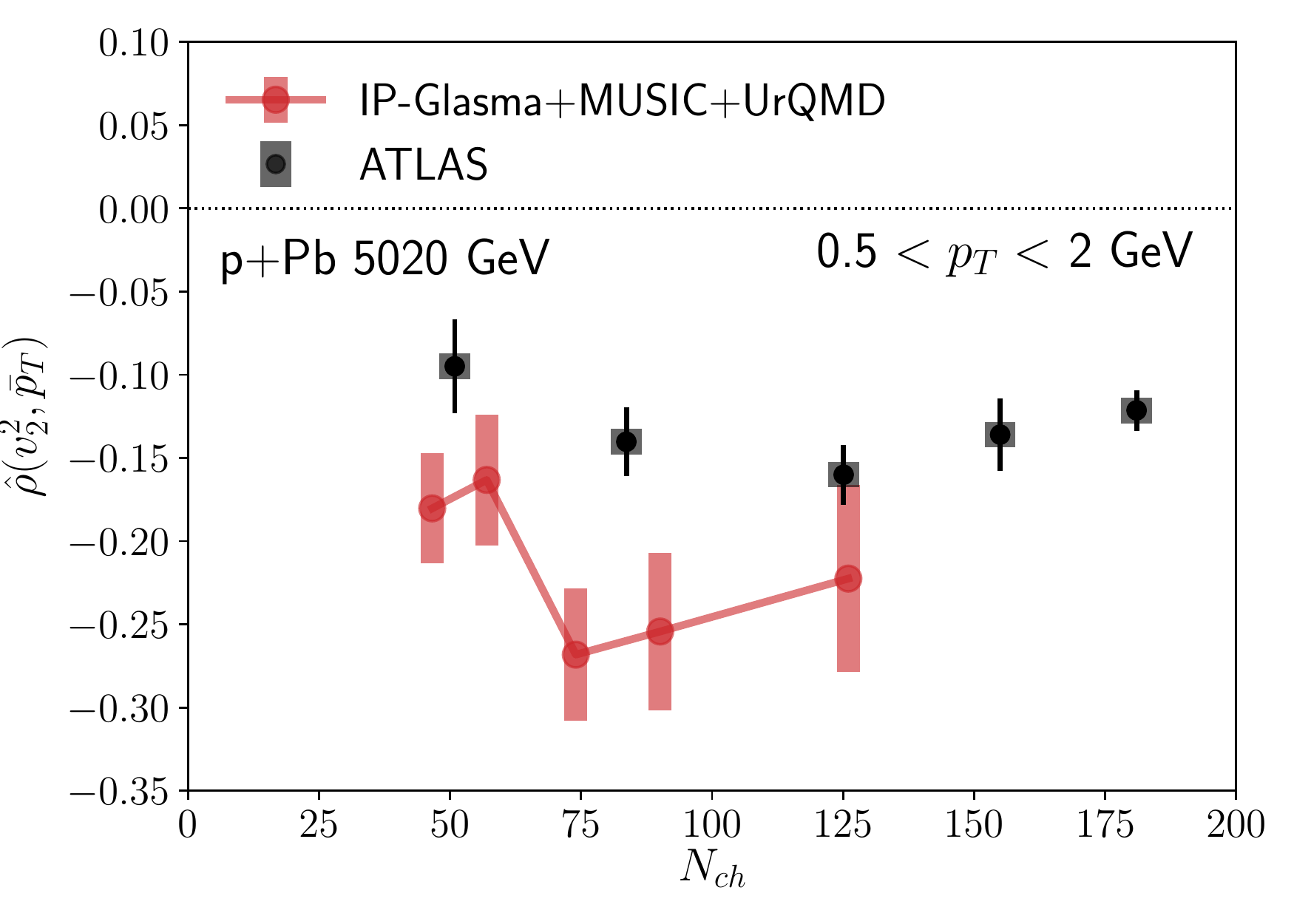}
\caption{The correlation measure between $v_2^2$ and mean transverse momentum $\hat{\rho}(v_2^2,\bar{p}_T)$ as a function of multiplicity compared to experimental data from the ATLAS Collaboration \cite{Aad:2019fgl}.  \label{fig:ATLAS-rho-corr-pT-pPb5020} } 
\end{figure}

We present results for $\hat{\rho}(v_2^2,\bar{p}_T)$ in p+Pb collisions at
$\sqrt{s}=5020\,{\rm GeV}$ in Fig.\,\ref{fig:ATLAS-rho-corr-pT-pPb5020}, and
compare to experimental data from the ATLAS Collaboration \cite{Aad:2019fgl}.
In this small system, the correlation is negative for all multiplicities, as it
is for the most peripheral Pb+Pb collisions. While our statistical errors are
large, we consistently underestimate the experimental data. The reason for the
disagreement could be non-flow contributions to the experimental data or
shortcomings in our model, such as the details of the initial state. Indeed, the
estimator $\hat{\rho}_{\rm est}$ (not shown) is even lower than the result for
$\hat{\rho}$, with $\hat \rho_{\rm est} \sim -0.5$ to $-0.4$. In Refs.~\cite{Bozek:2016yoj,Bozek:2020drh} it was found that a smoother, more Glauber like, initial state leads to a less negative (or even positive) $\hat \rho$ correlator.

\subsection{Effect of a pre-equilibrium stage: IP-Glasma+KoMPoST+MUSIC+UrQMD}\label{sec:kompost2}

\begin{figure}[tb]
\includegraphics[width=0.48\textwidth]{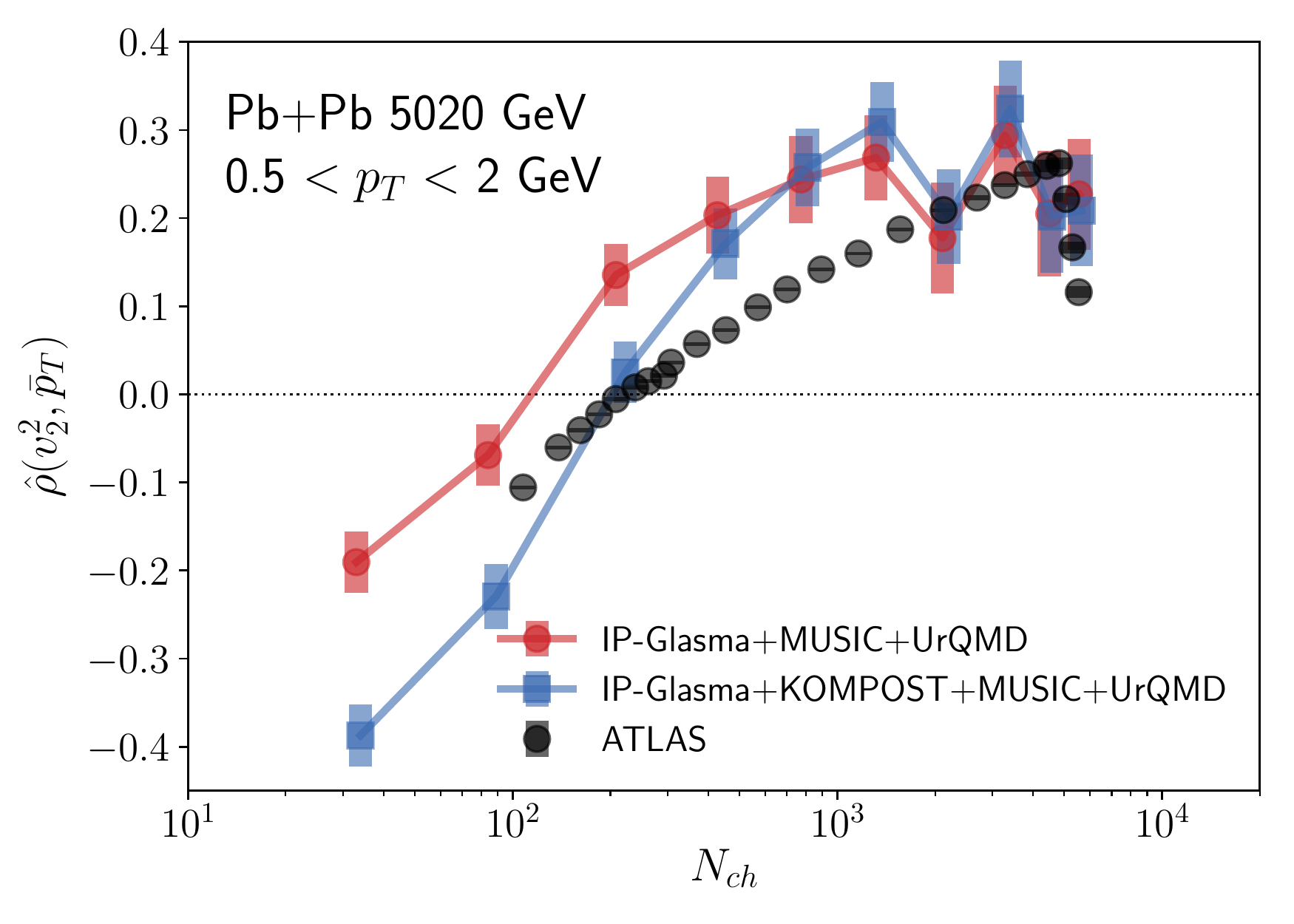}
  \caption{The correlation measure between $v_2^2$ and mean transverse momentum $\hat{\rho}(v_2^2,\bar{p}_T)$ as a function of multiplicity without (circles) and with (squares) a KoMPoST pre-equilibrium stage in the simulation. We compare to experimental data from the ATLAS Collaboration \cite{Aad:2019fgl} (black circles). \label{fig:ATLAS-rho-corr-pT-PbPb5020-KOMPOST}}
\end{figure}

As previously done for the mean transverse momentum fluctuation measure $\sqrt{C_m}/\langle\bar{p}_T\rangle$, here we study the effect of the pre-equilibrium KoMPoST stage on the $v_2^2$-$\bar{p}_T$ correlator $\hat{\rho}(v_2^2,\bar{p}_T)$. 
Fig.\,\ref{fig:ATLAS-rho-corr-pT-PbPb5020-KOMPOST} shows the comparison of the previously shown IP-Glasma+\textsc{Music}+UrQMD result to the one including a KoMPoST stage and experimental data from the ATLAS Collaboration. Because, again, differences between the simulation with and without KoMPoST are largest at small multiplicities, we use a logarithmic scale for $N_{\rm ch}$ on the $x$-axis.

For the elliptic flow - mean transverse momentum correlation $\hat{\rho}(v_2^2,\bar{p}_T)$, the KoMPoST stage leads to a reduction for the ATLAS $N_{\rm ch} \lesssim 400$. It is expected that the effect of the pre-equilibrium stage is larger for smaller multiplicities, where the hydrodynamic evolution is shorter. Both simulations, with and without KoMPoST, overestimate the experimental data in the range $200 \lesssim N_{\rm ch} \lesssim 2000$.

\section{Conclusions}\label{sec:conclusions}
Building on earlier work~\cite{Olszewski:2017vyg}, in \Sect{sec:v0defs} we introduced an integrated measure of transverse momentum fluctuations at fixed multiplicity, $v_0 \equiv \hat{\sigma}_{P_T}/\langle P_T \rangle$,
and the corresponding differential quantity $v_0(p_T)$. Both of these quantities can be measured using two-particle
correlations, provided these correlations factorize.  The observable $v_0(p_T)$ is analogous to the anisotropic flow
coefficients $v_n(p_T)$ for $n>0$, and its measurement can and should be performed
using the standard techniques to eliminate non-flow.

Within a hybrid model consisting of the IP-Glasma initial state, \textsc{Music}
hydrodynamics, and UrQMD hadronic cascade, we confirmed the factorization of
the two-particle correlation function into a product of $v_0(p_T^a)$ and
$v_0(p_T^b)$ (\Fig{fig:C-PbPb-C30-40-5020GeV-PID}). We further compared the model calculation to the simple model of
\Ref{Gardim:2019iah}, that assumes an exponential spectrum whose fluctuations
are determined by fluctuations of the mean transverse momentum and the
multiplicity.

We further made predictions for $v_0(p_T)$ of charged hadrons and identified
particles ($\pi,K,p)$, within the hybrid model in
$\sqrt{s}=5020\,{\rm GeV}$ Pb+Pb collisions (\Fig{fig:v0-PbPb-C30-40-5020GeV-PID}). 
The splitting between the  hadron species is a telltale signature of radial flow fluctuations.
The signal is rather strong, and it should be straight forward to perform the measurement. We further demonstrated that the centrality dependence of charged hadron $v_0(p_T)$ is dominated by the scaling of fluctuations with $(dN_{\rm ch}/d\eta)^{-1/2}$.

The $p_T$ integrated $v_0$ is closely related to previously measured mean transverse momentum fluctuation observables, at least when measured in narrow bins of multiplicity.
The partial
correlation method allows for larger multiplicity bins with a corresponding increase in statistics. We discussed the relation of $v_0$ to observables previously measured by the ALICE and ATLAS Collaborations in \Sect{sect:relationto} and presented results for those observables obtained in our hybrid model in \Sect{Sect:ComparisonWpt}. These measurements suffer from non-flow correlations and should be repeated before they can be fairly compared to hydrodynamic simulations.

The magnitude of $v_0$ gives direct information on the fluctuations in the initial
state.  Indeed, we found that the entropy per area  is
an excellent $v_0$ estimator, with a correlation coefficient of  90-95\%  
over a large part of the centrality range (\Fig{fig:Q-differentEstimators-PbPb5020}).\footnote{For very peripheral systems, the average event entropy density is a better estimator as the interaction area is likely composed of disconnected regions.} Thus measurements of $v_0$ can tightly constrain the fluctuations in the system size at fixed initial entropy.
Interestingly, the estimator gave a somewhat better
description of the centrality dependence of the ALICE data on $p_T$ fluctuations than the full hybrid model result (\Fig{fig:CmOverMpT-PbPb-2760GeV} and \Fig{fig:CmOverMpT-PbPb-2760GeV-Estimator}). This might indicate that final state parameters, such as the
temperature dependent bulk viscosity, were not chosen correctly. In fact, we
demonstrated that the mean transverse momentum fluctuations are sensitive to
the choice of transport parameters (\Fig{fig:CmOverMpT-PbPb-2760GeV-shearonly}). This lets us conclude that $v_0$ should be
included in every effort to constrain the transport coefficients of QCD using
heavy ion collision data.

We further studied the sensitivity of transverse momentum fluctuation measures
on the pre-equilibrium stage of the collision by comparing to simulations that
transition from the IP-Glasma Yang-Mills stage to hydrodynamics via an
off-equilibrium evolution provided by the KoMPoST model (\Fig{fig:CmOverMpT-PbPb-2760GeV-KOMPOST}). We found that
differences from the usual hybrid model were significant for multiplicities
$dN_{\rm ch}/d\eta \lesssim 200$, demonstrating the power of such observables to access
the early time dynamics for small (low multiplicity) systems.

In addition to the analysis of transverse momentum fluctuations, we studied the
correlation of elliptic flow with the event-by-event mean transverse momentum
within our hybrid model, and compared to experimental data from the ATLAS
Collaboration (\Fig{fig:ATLAS-rho-corr-PbPb5020}). We reproduced the qualitative features of the data, including a
sign change as a function of multiplicity, and used the 
entropy per area estimator to explain the simulation results with geometrical 
reasoning.

The quantitative agreement of our result with the experimental data is not
perfect, especially at smaller multiplicities, which could be a result of
non-flow in the experimental data, or shortcomings of the model such as the use
of less than optimal parameters. Potential problems of non-flow are
particularly important for small systems. In p+Pb collisions, our model
overpredicts the strength of the elliptic flow - mean transverse momentum correlation for most
multiplicity bins (\Fig{fig:ATLAS-rho-corr-pT-pPb5020}). We therefore urge the experiments to reduce the non-flow
contributions by employing similar methods for the mean transverse momentum
measure as for anisotropic flow coefficients.

To understand the influence of the thermalization stage on the $v_2^2$-$\bar{p}_T$ correlations,  we again compared the
simulation results to the experimental data, with and without the KoMPoST thermalization module (\Fig{fig:ATLAS-rho-corr-pT-PbPb5020-KOMPOST}). Again we find that the thermalization stage has a significant effect on this observable at small multiplicities,  $dN_{\rm ch}/d\eta \lesssim 200$. 

In the future, measurements of $v_0(p_T)$ can be used to diagnose the quark gluon plasma in different regimes. For instance, since $v_0$ is tightly correlated with the entropy density at fixed system size,  $v_0(p_T)$ could be measured for jets and others penetrating probes, providing new constraints on the temperature dependence of energy loss.  
Similarly, measurements of $v_0(p_T)$ could clarify the ``no-man's land'', a
region of momentum from $2 \ldots 6\,{\rm GeV}$, which reflects the
transition between hydrodynamics to jet-quenching. 
An upward fluctuation in the entropy (at fixed system size) leads to more hydrodynamic particles at a given $p_T$, and simultaneously suppresses the jet contribution through additional energy-loss.
Thus, there could be a sign change in $v_0(p_T)$ depending on the details of the jet energy loss model.
Finally, $v_0(p_T)$ can provide important constraints on the initial state fluctuations of the entropy per area in $p+A$ and $p+p$ collisions, and thus help to clarify the thermalization dynamics~\cite{Kurkela:2019kip}.   We hope that the current manuscript can motivate and guide both theorists and experimentalists in these next steps.

\section*{Acknowledgments} 
We thank Piotr Bozek, Giuliano Giacalone, and Aleksas Mazeliauskas for useful discussions.
BPS is supported under DOE Contract No. DE-SC0012704. CS is supported under DOE Contract No. DE-SC0013460.
DT is supported under DOE Contract No. DE\nobreakdash-FG\nobreakdash-02\nobreakdash-08ER41450.
This work is in part supported by the U.S. Department of Energy, Office of Science, Office of Nuclear Physics, within the framework of the Beam Energy Scan Theory (BEST) Topical Collaboration. This research used resources of the National Energy Research Scientific Computing Center, which is supported by the Office of Science of the U.S. Department of Energy under Contract No. DE-AC02-05CH11231 and resources of the high performance computing services at Wayne State University.


%\bibliography{spires}

\end{document}